\documentclass[
reprint,
amsmath,
amssymb,
aps,
superscriptaddress,
notitlepage
]{revtex4-1}

\usepackage[colorlinks, breaklinks=true, linkcolor=blue, citecolor=blue, linktocpage=true]{hyperref}
\usepackage{color}
\usepackage{graphicx}
\usepackage{dcolumn}
\usepackage{bm}

\usepackage{txfonts}

\begin{document}

\title{Quantum Kinetic Theory of Thermoelectric and Thermal Transport in a Magnetic Field}

\author{Akihiko Sekine}
\email{akihiko.sekine@riken.jp}
\affiliation{RIKEN Center for Emergent Matter Science (CEMS), Wako, Saitama 351-0198, Japan}
\author{Naoto Nagaosa}
\affiliation{RIKEN Center for Emergent Matter Science (CEMS), Wako, Saitama 351-0198, Japan}
\affiliation{Department of Applied Physics, The University of Tokyo, Bunkyo, Tokyo 113-8656, Japan}
\date{\today}

\begin{abstract}
We present a general quantum kinetic theory that accounts for the interplay between a temperature gradient, momentum-space Berry curvatures of Bloch electrons, and Bloch-state scattering.
Using a theory that incorporates the presence of a temperature gradient by introducing a ``thermal vector potential'', we derive a quantum kinetic equation for Bloch electrons in the presence of disorder and a temperature gradient.
In contrast to the semiclassical Boltzmann formalism in which a temperature gradient is introduced by setting $\dot{\bm{r}}\cdot\frac{\partial}{\partial\bm{r}}\to \dot{\bm{r}}\cdot\frac{\partial T}{\partial\bm{r}}\frac{\partial}{\partial T}$ in the Boltzmann equation, the presence of a temperature gradient in our formalism is described as a driving force just as in the case of an electric field (i.e., comes from $\dot{\bm{k}}$ in the language of the semiclassical Boltzmann formalism).
Taking also into account the presence of electric and magnetic fields, the quantum kinetic equation we derive makes it possible to compute transport coefficients at arbitrary orders of electric-field $\bm{E}$, magnetic-field $\bm{B}$, and temperature-gradient $\nabla T$ strengths $|\bm{E}|^a |\bm{B}|^b |\nabla T|^c$.
Our theory enables a systematic calculation of magnetothermoelectric and magnetothermal conductivities of systems with momentum-space Berry curvatures.
As an illustration, we derive from a general microscopic electron model a general expression for the rate of pumping of electrons between valleys in parallel temperature gradient and magnetic field.
From this expression we find a relation, which is analogous to the Mott relation, between the rate of pumping due to a temperature gradient and that due to an electric field.
We also apply our theory to a two-band model for Weyl semimetals to study thermoelectric and thermal transport in a magnetic field.
We show that the Mott relation is satisfied in the chiral-anomaly induced thermoelectric conductivity, and that the Wiedemann-Franz law is violated in the chiral-anomaly induced thermal conductivity, which are both consistent with the results obtained by invoking semiclassical wave-packet dynamics.
\\
\end{abstract}

\maketitle

\section{Introduction}
Momentum-space Berry curvatures of Bloch electrons, which can be nonzero in systems with broken time-reversal symmetry and/or broken inversion symmetry \cite{Xiao2010}, have been revealed to play important roles in electronic transport phenomena.
An important and well recognized example is the anomalous Hall effect \cite{Nagaosa2010}, the Hall effect at zero magnetic field in systems with broken time-reversal symmetry.
The anomalous Hall conductivity of massive Dirac fermions on the surface of three-dimensional (3D) topological insulators is independent of disorder scattering and characterized completely by the momentum-space Berry curvature which gives rise to a nonzero Chern number, when the Fermi level is inside the gap \cite{Qi2008,Hasan2010,Qi2011,Nomura2011,Chang2013}.
More recently, it was found that nonlinear (second-order) anomalous Hall effect can be realized due to the Berry curvature dipole even in systems with time-reversal symmetry \cite{Sodemann2015,Du2018,Zhang2018,You2018,Ma2019,Kang2019}.
Another example is the negative magnetoresistance in 3D Weyl and Dirac semimetals \cite{Son2013,Burkov2014,Xiong2015,Li2015,Huang2015,Sekine2017}, which arises as a consequence of a condensed-matter realization of the chiral anomaly, i.e., the pumping of electrons between valleys in parallel electric and magnetic fields.
The effects of nontrivial band structures giving rise to nonzero Berry curvatures on electronic properties are often referred to as the Berry phase effects \cite{Xiao2010}, and have been studied intensively and extensively in condensed matter physics.

As well as an electric field, a temperature gradient can also drive electrons into non-equilibrium states, resulting in a generation of currents such as electric current.
Studies of transport induced by a temperature gradient have a long history and are one of the important subjects in condensed matter physics.
Among them, spin transport phenomena induced by a temperature gradient, often called the spin caloritonics \cite{Bauer2012}, is an emerging field in this decade, triggered by the discovery of the spin Seebeck effect \cite{Uchida2008}.
The Berry phase effects on transport induced by a temperature gradient are often taken into account by combining semiclassical wave-packet dynamics with semiclassical Boltzmann theory \cite{Xiao2010}, as in the case of transport induced by an electric field.
In this phenomenological method, the calculation becomes complicated when a magnetic field is present \cite{Kim2014a,Lundgren2014,Sharma2016,Nandy2019}: the form of the correction to the distribution function is assumed and then its solution is obtained by substituting the assumed form into the Boltzmann equation.
This study is motivated by the need for a more straightforward and practical theory that is applicable to realistic models.

In this paper, we develop a quantum kinetic theory for electronic transport induced by a temperature gradient in a magnetic field in weakly disordered systems with large momentum-space Berry curvatures, which fully accounts for the interplay between the presence of temperature gradient and electric and magnetic fields, momentum-space Berry curvatures, and Bloch-state scattering.
We employ a recent theory \cite{Tatara2015} that proposes that a temperature gradient can be described by a ``thermal vector potential'' in analogy with electromagnetic vector potential.
We take the effect of magnetic fields into account using a semiclassical approximation that we expect to be accurate when the magnetic field is weak enough that Landau quantization can be neglected.
We also take the effect of disorder into account within the Born approximation.
Our theory enables a systematic calculation of the density matrix induced by a temperature gradient in powers of the magnetic field strength, from which physical observables such as thermoelectric and thermal conductivities can be obtained.

This paper is organized as follows.
In Sec.~\ref{Sec-QKE} we derive a quantum kinetic equation for Bloch electrons in the presence of a temperature gradient, disorder, and electric and magnetic fields by applying a Wigner transformation to the quantum Liouville equation.
In Sec.~\ref{Sec-Density-matrix} we obtain the solution of the density matrix induced by a temperature gradient in the absence of electric and magnetic fields, and then obtain a general expression for the density matrix in a magnetic field by performing a low-magnetic-field expansion.
In Sec.~\ref{Sec-Transport-coefficients}, to check the validity of our theory, we derive general expressions for the intrinsic (i.e., Berry phase) contribution to the Nernst effect at zero magnetic field and its reciprocal effect, for which we find that the Onsager reciprocal relation is indeed satisfied.
In Sec.~\ref{Sec-CA-generic} we apply our theory to a genetic microscopic electron model and derive a general expression for the rate of pumping of electrons between valleys (i.e., the thermal chiral anomaly) in parallel temperature gradient and magnetic field.
In Sec.~\ref{Sec-WSM} we apply our theory to a simple two-band model of Weyl semimetals and demonstrate that the Mott relation is satisfied for the chiral-anomaly induced magnetothermoelectric conductivity, while the Wiedemann-Franz law is violated for the chiral-anomaly induced magnetothermal conductivity.
In Sec.~\ref{Sec-Discussion} we discuss our theory in connection with possible applications of our theory.
Finally, in Sec.~\ref{Sec-Summary} we summarize this study.

\section{Quantum kinetic equation \label{Sec-QKE}}
In this section, we derive a quantum kinetic equation for Bloch electrons in the presence of disorder, a temperature gradient, and electric and magnetic fields.
Throughout this paper, we work in the basis of the disorder-free Hamiltonian eigenstates, which we refer to as the eigenstate basis:
\begin{align}
H_0 |m,\bm{k}\rangle=\varepsilon^m_{\bm{k}} |m,\bm{k}\rangle,
\end{align}
where $H_0$ is the crystal Hamiltonian, $\varepsilon^m_{\bm{k}}$ is an eigenvalue of $H_0$, $\bm{k}$ is a momentum in the crystal's Brillouin zone, and $m$ is a band index.
We consider a generic single-particle $D$-dimensional Bloch Hamiltonian $H_0(\bm{p})$ with momentum operator $\bm{p}=-i\hbar\nabla$.
In the presence of an electromagnetic vector potential $\bm{A}(\bm{r},t)$, minimal coupling results in $\bm{p}\rightarrow \bm{p}+e\bm{A}$.
Throughout this paper, we adopt the notation $e>0$ which is the magnitude of the electron charge.

\subsection{Introducing disorder}
In the absence of external fields, the total Hamiltonian of the system is $H=H_0+U$ where $U$ is the disorder potential.
We treat disorder effects on transport coefficients within the Born approximation.
Our starting point is the quantum kinetic equation in the presence of disorder \cite{Culcer2017}
\begin{align}
\frac{\partial \langle\rho\rangle}{\partial t}+\frac{i}{\hbar}[H_0,\langle\rho\rangle]+K(\langle\rho\rangle)=0,
\label{QLE-Born}
\end{align}
where $\langle\rho\rangle$ is the disorder-averaged density-matrix operator of the system and the scattering term $K(\langle\rho\rangle)$ is given by
\begin{align}
K(\langle\rho\rangle)=\frac{1}{\hbar^2}\int_0^\infty dt'\, \left\langle \left[U, [e^{-iH_0t'/\hbar}U e^{iH_0t'/\hbar}, \langle\rho(t)\rangle]\right]\right\rangle,
\label{scattering-K}
\end{align}
where $\langle\rho(t)\rangle=e^{-iH_0 t/\hbar}\langle\rho\rangle e^{iH_0 t/\hbar}$.
The full expression for $K(\langle\rho\rangle)$ can be found in Ref.~\cite{Culcer2017}.
We separate the density matrix $\langle\rho\rangle$ in the eigenstate basis into the band-diagonal part $\langle n\rangle$ and the band-off-diagonal part $\langle S\rangle$ using the notation $\langle\rho\rangle=\langle n\rangle+\langle S\rangle$.
Then, the scattering term $K(\langle\rho\rangle)$ can be separated into four parts which map $\langle n\rangle$ and $\langle S\rangle$ to band-diagonal and band off-diagonal contributions to $\partial \langle\rho(t)\rangle/\partial t$ in Eq.~(\ref{QLE-Born}).
To leading order in disorder strength, the contribution from $\langle n\rangle$, i.e., $K(\langle n\rangle)$, becomes dominant in many cases.
In the case of elastic scattering the band-diagonal part of $K(\langle n\rangle)$ $[\equiv I(\langle n\rangle)]$ is given by  \cite{Culcer2017}
\begin{align}
[I(\langle n\rangle)]^{mm}_{\bm{k}} =\frac{2\pi}{\hbar} \sum_{m',\bm{k}'} & \langle U^{mm'}_{\bm{k}\bm{k}'}U^{m'm}_{\bm{k}'\bm{k}}\rangle(n^{m}_{\bm{k}} - n^{m'}_{\bm{k}'})\delta(\varepsilon^m_{\bm{k}} - \varepsilon^{m'}_{\bm{k}'}),
\label{scattering-I}
\end{align}
with $m$ and $m'$ being band indices.
This is exactly Fermi's golden rule.
Similarly, the band-off-diagonal part of $K(\langle n\rangle)$ $[\equiv J(\langle n\rangle)]$ is given by \cite{Culcer2017}
\begin{align}
[J(\langle n\rangle)]^{mm''}_{\bm{k}} =&\ \frac{\pi}{\hbar} \sum_{m',\bm{k}'} \langle U^{mm'}_{\bm{k}\bm{k}'}U^{m'm''}_{\bm{k}'\bm{k}}\rangle \left[ (n^{m}_{\bm{k}} - n^{m'}_{\bm{k}'}) \right. \nonumber\\
&\left.\times\ \delta(\varepsilon^m_{\bm{k}} - \varepsilon^{m'}_{\bm{k}'}) +(n^{m''}_{\bm{k}} - n^{m'}_{\bm{k}'}) \delta(\varepsilon^{m''}_{\bm{k}} - \varepsilon^{m'}_{\bm{k}'})\right], 
\label{scattering-J}
\end{align}
where  $m \ne m''$.
The contribution from this term corresponds to the vertex correction in the ladder-diagram approximation of perturbation theory \cite{Culcer2017,Sekine2017}.
See also Ref.~\cite{Xiao2017a} for the semiclassical derivation of the term corresponding to Eq.~(\ref{scattering-J}).
Equivalently, in the context of the anomalous Hall effect \cite{Nagaosa2010,Sinitsyn2007,Sinitsyn2008,Xiao2017}, the contribution from Eq.~(\ref{scattering-J}) describes the side-jump velocity contribution \cite{Nandy2019a}.
However, in the case of 2D massive Dirac model \cite{Nandy2019a,Xiao2019}, the contribution from Eq.~(\ref{scattering-J}) gives rise to only one half of the total side-jump contribution defined in Ref.~\cite{Sinitsyn2007}.

Here, we note that the contribution from the band off-diagonal part of the density matrix $\langle S\rangle$ to the scattering term, i.e., $K(\langle S\rangle)$, was studied recently in detail in the second-order nonlinear Hall effect \cite{Nandy2019a}, and it was shown that the contribution from the diagonal part of $K(\langle S\rangle)$ $[\equiv I(\langle S\rangle)]$, which is given by
\begin{align}
[I(\langle S\rangle)]^{mm}_{\bm{k}} =&\ \frac{\pi}{\hbar} \sum_{m',m'',\bm{k}'}\left[ \langle U^{mm'}_{\bm{k}\bm{k}'}U^{m'm''}_{\bm{k}'\bm{k}}\rangle\langle S\rangle_{\bm{k}}^{m''m}\delta(\varepsilon^{m'}_{\bm{k}'} - \varepsilon^{m''}_{\bm{k}})\right.\nonumber\\
&+\langle U^{m'm''}_{\bm{k}\bm{k}'}U^{m''m}_{\bm{k}'\bm{k}}\rangle\langle S\rangle_{\bm{k}}^{mm'}\delta(\varepsilon^{m'}_{\bm{k}} - \varepsilon^{m''}_{\bm{k}'})\nonumber\\
&-\langle U^{mm'}_{\bm{k}\bm{k}'}U^{m''m}_{\bm{k}'\bm{k}}\rangle\langle S\rangle_{\bm{k}'}^{m'm''}\delta(\varepsilon^{m''}_{\bm{k}'} - \varepsilon^{m}_{\bm{k}})\nonumber\\
&\left. -\ \langle U^{mm'}_{\bm{k}\bm{k}'}U^{m''m}_{\bm{k}'\bm{k}}\rangle\langle S\rangle_{\bm{k}'}^{m'm''}\delta(\varepsilon^{m}_{\bm{k}} - \varepsilon^{m'}_{\bm{k}'})\right], 
\label{scattering-J_d-S}
\end{align}
describes the skew scattering contribution to the Hall conductivity \cite{Nandy2019a}.
On the other hand, it was also pointed out recently that there exists a contribution arising from the interband-coherence effect of dc electric fields during scattering, which is specific to nonlinear response, i.e., has no counterpart in linear response \cite{Xiao2019}.

\subsection{Introducing a temperature gradient}
Now, we take the effects of a temperature gradient into account.
According to the famous Luttinger's proposal \cite{Luttinger1964}, a temperature gradient can be described by a scalar potential $\Psi$ satisfying $-\nabla\Psi=-\nabla T/T$.
In a similar context, Tatara recently introduced a ``thermal vector potential'' $\bm{A}_T$ that satisfies \cite{Tatara2015}
\begin{align}
-\frac{\partial \bm{A}_T}{\partial t}=-\frac{\nabla T}{T}\equiv \bm{E}_T.
\end{align}
In analogy with electromagnetism, a temperature gradient is also termed a ``gravitoelectric field'' $\bm{E}_T$ \cite{Mashhoon-Book,Ryu2012,Nomura2012,Sekine2016,Shitade2019}.
In Ref.~\cite{Tatara2015} it was suggested that, in single-band systems such as parabolic band system
with the energy $\varepsilon_{\bm{p}}=\hbar^2\bm{p}^2/2m$, the thermal vector potential is incorporated via minimal coupling form
\begin{align}
\bm{p}\to\bm{p}-\varepsilon_{\bm{p}}\bm{A}_T.
\label{thermal-minimal-coupling}
\end{align}
However, it is not trivial to generalize this incorporation of the thermal vector potential in the case of multiband systems.

To resolve this difficulty, we here propose to generalize the Wigner distribution function in the presence of an electromagnetic vector potential \cite{Vasko-book,Sekine2017} to the form containing a thermal vector potential $\bm{A}_T$ as
\begin{align}
\langle\rho\rangle_{\bm{p}}^{mn}(\bm{r})=\frac{1}{2}\int d^D\bm{R}\, \langle m,\bm{r}_+|\left\{e^{-(i/\hbar)\bm{P}_T\cdot\bm{R}},\, \rho\right\}| n,\bm{r}_-\rangle,
\label{Wigner-distribution-thermal}
\end{align}
where $\{\ ,\ \}$ is the anticommutator for operators, $\bm{P}_T=\bm{p}+H_0\bm{A}_T$, $\bm{r}_\pm=\bm{r}\pm\bm{R}/2$, and $| n,\bm{r}\rangle=\sum_{m,\bm{k}} |m,\bm{k}\rangle\langle m,\bm{k}| n,\bm{r}\rangle=\sum_{\bm{k}}e^{i\bm{k}\cdot\bm{r}}|n,\bm{k}\rangle$ is the Fourier transform of $|n,\bm{k}\rangle$.
Note that the sign in front of $\mathcal{H}_0\bm{A}_T$ in $\bm{P}_T$ is different from the one in the minimal coupling~(\ref{thermal-minimal-coupling}), as in the case of electromagnetic vector potential \cite{Vasko-book,Sekine2017}.
We perform the generalized Wigner transformation~(\ref{Wigner-distribution-thermal}) on the first two terms in Eq.~(\ref{QLE-Born}) (i.e., the usual quantum Liouville equation $\partial \rho/\partial t+\frac{i}{\hbar}[H_0, \rho]=0$) as
\begin{align}
\int d^D\bm{R}\, \langle m,\bm{r}_+|\left\{e^{-(i/\hbar)\bm{P}_T\cdot\bm{R}},\, \left(\frac{\partial \rho}{\partial t}+\frac{i}{\hbar}[H_0,\rho]\right)\right\} | n,\bm{r}_-\rangle=0.
\label{Wigner-trans-QLE}
\end{align}
Note that the scattering term $K(\langle\rho\rangle)$ is not changed after the Wigner transformation.
We use the identities
\begin{align}
e^{-(i/\hbar)\bm{P}_T\cdot\bm{R}}\, \frac{\partial\rho}{\partial t}=\frac{\partial}{\partial t}\left(e^{-(i/\hbar)\bm{P}_T\cdot\bm{R}}\, \rho\right)+\frac{i}{\hbar}\frac{\nabla T}{T}\cdot\bm{R}H_0 e^{-(i/\hbar)\bm{P}_T\cdot\bm{R}}\, \rho,\nonumber\\
\frac{\partial\rho}{\partial t}\, e^{-(i/\hbar)\bm{P}_T\cdot\bm{R}}=\frac{\partial}{\partial t}\left(\rho\, e^{-(i/\hbar)\bm{P}_T\cdot\bm{R}}\right)+\frac{i}{\hbar}\rho\, e^{-(i/\hbar)\bm{P}_T\cdot\bm{R}}\frac{\nabla T}{T}\cdot\bm{R}H_0,
\label{time-derivative-identities}
\end{align}
where $-\partial \bm{A}_T/\partial t=-\nabla T/T$ is the temperature gradient.
Note that the positions of $\bm{R}$ on the right-hand sides of Eq.~(\ref{time-derivative-identities}) are arbitrary, since $\bm{R}$ is just a number before it is sandwiched by the eigenstate basis.
Inserting the completeness relation $\sum_n\int d\bm{r}\, |n, \bm{r}\rangle \langle n, \bm{r}|=1$, using the Fourier transform $| n,\bm{r}\rangle=\sum_{\bm{k}}e^{i\bm{k}\cdot\bm{r}}|n,\bm{k}\rangle$, and noting that $\bm{R}=i\hbar\nabla_{\bm{p}}$ from the definition of the Wigner distribution function~(\ref{Wigner-distribution-thermal}), we obtain one of the matrix  elements in Eq.~(\ref{Wigner-trans-QLE}) as
\begin{widetext}
\begin{align}
\int d^D\bm{R}\, \langle m,\bm{r}_+|\bm{R}H_0e^{-(i/\hbar)\bm{P}_T\cdot\bm{R}}\rho| n,\bm{r}_-\rangle
&=\sum_{n',n''}\langle m,\bm{p}|i\hbar\nabla_{\bm{p}}|n', \bm{p}\rangle \langle n', \bm{p}|H_0|n'', \bm{p}\rangle\langle\rho\rangle_{1\bm{p}}^{n'' n}\nonumber \\
&=i\hbar\left[\mathcal{H}_{0\bm{p}}^{mn''}\nabla_{\bm{p}} \langle\rho\rangle_{1\bm{p}}^{n''n}+\langle u^{m}_{\bm{p}}|\nabla_{\bm{p}}u^{n'}_{\bm{p}}\rangle\mathcal{H}_{0\bm{p}}^{n'n''}\langle\rho\rangle_{1\bm{p}}^{n''n}-\mathcal{H}_{0\bm{p}}^{mn'}\langle\rho\rangle_{1\bm{p}}^{n'n''}\langle u^{n''}_{\bm{p}}|\nabla_{\bm{p}}u^{n}_{\bm{p}}\rangle\right],
\label{matrix-element}
\end{align}
\end{widetext}
where $\langle\rho\rangle_{1\bm{p}}^{n'' n}=\int d^D\bm{R}\, \langle m,\bm{r}_+|e^{-(i/\hbar)\bm{P}_T\cdot\bm{R}}\rho| n,\bm{r}_-\rangle$, $\mathcal{H}_{0\bm{p}}^{n'n''}=\langle n', \bm{p}|H_0|n'', \bm{p}\rangle=\delta_{n'n''}\varepsilon_{\bm{p}}^{n'}$, and the third term on the right-hand side comes from the Hermiticity of the equation and 
consistency with the single-band limit \cite{Sekine2017}.
Note that $\nabla_{\bm{p}}$ in the first term on the right-hand side acts only on the distribution function, since our formalism recovers the semiclassical Boltzmann equation in the single-band limit \cite{Sekine2017}.
Other possible terms can be obtained in a similar way as Eq.~(\ref{matrix-element}).
Finally, after a calculation we arrive at the quantum kinetic equation in the presence of disorder and a temperature gradient $-\nabla T/T$:
\begin{align}
\frac{\partial \langle\rho\rangle}{\partial t}+\frac{i}{\hbar}[\mathcal{H}_0,\langle \rho\rangle]+K(\langle\rho\rangle)=D_T(\langle\rho\rangle),
\label{kinetic-equation-thermal}
\end{align}
where $\bm{k}=\bm{p}/\hbar$ is the crystal wave vector and $D_{T}(\langle\rho\rangle)$ is the thermal driving term given by
\begin{align}
D_{T}(\langle\rho\rangle)=\frac{1}{2\hbar}\frac{\nabla T}{T}\frac{D(\{\mathcal{H}_0, \langle\rho\rangle\})}{D\bm{k}}\label{driving-term-T},
\end{align}
with $\{ \ ,\ \}$ being an anticommutator.
Here, $D/D\bm{k}$ is the covariant derivative acting on matrices defined by
\begin{align}
\frac{D X}{D\bm{k}}=\nabla_{\bm{k}}X-i[\bm{\mathcal{R}}_{\bm{k}}, X],
\label{Covariant-derivative}
\end{align}
where $X$ is a matrix, $[\ ,\ ]$ is a commutator, and $\bm{\mathcal{R}}_{\bm{k}}=\sum_{a=x,y,z}\mathcal{R}_{\bm{k}}^a\bm{e}_a$ with $[\mathcal{R}_{\bm{k}}^a]^{mn}=i\langle u^m_{\bm{k}}|\partial_{k_a}u^n_{\bm{k}}\rangle$ being the generalized Berry connection.
Note that, the usual derivative $\nabla_{\bm{k}}$ in Eq.~(\ref{driving-term-T}) acts only on $\langle \rho\rangle$, although we have introduced a covariant derivative notation in Eq.~(\ref{driving-term-T}) to make the notation consistent with the electric and magnetic driving terms.

\subsection{Including electric and magnetic fields}
Next, we take the effects of electric and magnetic fields into account using a semiclassical approximation that we expect to be accurate when the weak magnetic field condition $\omega_{\mathrm{c}}\tau\ll 1$ is satisfied and Landau quantization can be neglected.
Here, $\omega_{\mathrm{c}}$ is the cyclotron frequency and $\tau$ is the transport relaxation time.
Note that there is no limitation of the strength of the electric field.
Inclusion of electric and magnetic fields into the kinetic equation~(\ref{kinetic-equation-thermal}) is easily done by adding an electromagnetic vector potential $\bm{A}$ to $\bm{P}_T$ in Eq.~(\ref{Wigner-distribution-thermal}).
Namely, we set $\bm{P}_T=\bm{p}+H_0\bm{A}_T-e\bm{A}$, where $e>0$ is the magnitude of the electron charge.
The resulting electric and magnetic fields reads $\bm{E}=-\partial \bm{A}/\partial t$ and $\bm{B}=\nabla\times\bm{A}$, respectively.
The electric- and magnetic-field dependent terms come respectively from the Wigner transformation on $\partial\rho/\partial t$ and $[H_0,\rho]$ \cite{Sekine2017}, which means that we can obtain such terms separately from the thermal driving term~(\ref{driving-term-T}).
Then, combining Eq.~(\ref{kinetic-equation-thermal}) and the result obtained in Ref.~\cite{Sekine2017}, we arrive at the quantum kinetic equation in the presence of disorder, an electric field $\bm{E}$, a magnetic field $\bm{B}$, and a temperature gradient $-\nabla T/T$:
\begin{align}
&\frac{\partial \langle \rho\rangle}{\partial t}+\frac{i}{\hbar}[\mathcal{H}_0,\langle \rho\rangle]+\frac{1}{2\hbar}\left\{\frac{D\mathcal{H}_0}{D\bm{k}} \cdot \nabla\langle \rho\rangle\right\}+K(\langle \rho\rangle)\nonumber\\
&=D_E(\langle \rho\rangle)+D_B(\langle \rho\rangle)+D_T(\langle \rho\rangle).
\label{full-kinetic-equation}
\end{align}
Here and below $\{\bm{a}\cdot\bm{b}\}\equiv\bm{a}\cdot\bm{b}+\bm{b}\cdot\bm{a}$ (with $\bm{a}$ and $\bm{b}$ being vectors) denotes a symmetrized operator product.
In Eq.~(\ref{full-kinetic-equation}) $D_E(\langle \rho\rangle)$ and $D_B(\langle \rho\rangle)$ are the electric and magnetic driving terms given by \cite{Sekine2017}
\begin{align}
D_{E}(\langle \rho\rangle)&=\frac{e\bm{E}}{\hbar}\cdot\frac{D\langle \rho\rangle}{D\bm{k}},\label{driving-term-E}
\end{align}
\begin{align}
D_{B}(\langle \rho\rangle)&=\frac{e}{2\hbar^2}\left\{\left(\frac{D \mathcal{H}_0}{D\bm{k}}\times\bm{B}\right)\cdot\frac{D\langle \rho\rangle}{D\bm{k}}\right\},\label{driving-term-B}
\end{align}
where is $D/D\bm{k}$ the covariant derivative defined in Eq.~(\ref{Covariant-derivative}).

The covariant derivatives reduce to simple derivatives in a spin-independent single-band system, for example in a parabolic band system with $\mathcal{H}_0({\bm{k}})=\hbar^2\bm{k}^2/2m\, (\equiv \varepsilon_{\bm{k}})$.
Accordingly, the quantum kinetic equation~(\ref{full-kinetic-equation}) reduces to
\begin{align}
&\left[\frac{\partial}{\partial t}+\bm{v}_{\bm{k}}\cdot\nabla-\frac{e}{\hbar}\left(\bm{E}+\bm{v}_{\bm{k}}\times\bm{B}\right)\cdot\nabla_{\bm{k}}-\frac{\varepsilon_{\bm{k}}-\mu}{\hbar T}\nabla T\cdot\nabla_{\bm{k}}\right]f_{\bm{k}}\nonumber\\
&=-I(f_{\bm{k}}),
\label{semiclassical-Boltzmann}
\end{align}
where we have defined the velocity $\bm{v}_{\bm{k}}=(1/\hbar)\nabla_{\bm{k}}\varepsilon_{\bm{k}}$ and $I(f_{\bm{k}})$ is the single-band version of Eq.~(\ref{scattering-I}).
Here, note that in the semiclassical Boltzmann equation the temperature gradient dependent term comes from the $\dot{\bm{r}}\cdot\nabla f$ term, which becomes $-\frac{\varepsilon_{\bm{k}}-\mu}{\hbar T}\nabla T\cdot\nabla_{\bm{k}}f$ in Eq.~(\ref{semiclassical-Boltzmann}) by using $\dot{\bm{r}}=\bm{v}_{\bm{k}}$ and $\frac{\partial f}{\partial T}=-\frac{\varepsilon_{\bm{k}}-\mu}{T}\frac{\partial f}{\partial \varepsilon_{\bm{k}}}$.
The quantum kinetic equation (\ref{full-kinetic-equation}) we have derived can therefore be understood as a generalization of the simple Boltzmann equation (\ref{semiclassical-Boltzmann}) in which the velocity and distribution function scalars are replaced by matrices, the simple derivatives $\nabla_{\bm{k}}$ are replaced by covariant derivatives $D/D\bm{k}$, and scalar products are replaced by symmetrized matrix products $\frac{1}{2}\{\ \ \cdot\ \ \}$.
Equations~(\ref{kinetic-equation-thermal}) and (\ref{full-kinetic-equation}) are the principal result of this paper.

\section{Solution of the density matrix \label{Sec-Density-matrix}}
In this section, by solving the quantum kinetic equation (\ref{full-kinetic-equation}), we give general expressions for the density matrix in the presence of electric and magnetic fields and a temperature gradient.
Especially, we obtain the linear response of the density matrix to a temperature gradient in a low magnetic field by performing the low-magnetic-field expansion.

\subsection{Density matrix at zero electric and magnetic fields \label{Sec-rho_T}}
Here, we consider electron transport induced solely by a temperature gradient, i.e., we set $\bm{E}=\bm{B}=0$ in Eq.~(\ref{full-kinetic-equation}).
We can follow the procedure that was done for the case of transport induced solely by an electric field \cite{Culcer2017,Sekine2017}.
In linear response, we write the electron density matrix $\langle\rho\rangle$ as $\langle\rho\rangle=\langle\rho_0\rangle+\langle\rho_T\rangle$, where $\langle\rho_0\rangle$ is the equilibrium density matrix and $\langle\rho_T\rangle$ is the correction to $\langle\rho_0\rangle$ which is linear in the temperature gradient $-\nabla T/T$.
With this notation we need to solve the kinetic equation in the form
\begin{align}
\frac{\partial \langle\rho_T\rangle}{\partial t}+\frac{i}{\hbar}[\mathcal{H}_0,\langle\rho_T\rangle]+K(\langle\rho_T\rangle)=D_T(\langle \rho_0\rangle),
\label{KE-for-rho_T}
\end{align}
where we have used that fact that $K(\langle\rho_0\rangle)=0$.
We divide the electron density matrix response $\langle\rho_T\rangle$ into the diagonal part $\langle n_T\rangle$ and the off-diagonal part $\langle S_T\rangle$, writing $\langle\rho_T\rangle=\langle n_T\rangle+\langle S_T\rangle$.
Note that the equilibrium density matrix $\langle\rho_0\rangle$ is diagonal in the band index.

When only band-diagonal to band-diagonal terms are included in the scattering kernel, it is easy to solve for the steady-state value of $\langle n_T\rangle$.
The kinetic equation (\ref{KE-for-rho_T}) in this limit is 
\begin{align}
[I(\langle n_T\rangle)]^{mm}_{\bm{k}}=[D_T(\langle \rho_0\rangle)]^{mm}_{\bm{k}}=\frac{\nabla T}{T}\cdot \bm{v}^m_{\bm{k}}\left(\varepsilon^m_{\bm{k}}-\mu\right)\frac{\partial f_0(\varepsilon^m_{\bm{k}})}{\partial \varepsilon^m_{\bm{k}}},
\end{align}
where $m$ is a band index, $\bm{v}^m_{\bm{k}}=(1/\hbar)\nabla_{\bm{k}}\varepsilon^m_{\bm{k}}$, and $\langle \rho_0\rangle^{mm}_{\bm{k}}=f_0(\varepsilon^m_{\bm{k}})$ is the Fermi-Dirac distribution function.
Here, $\varepsilon^m_{\bm{k}}$ is an eigenvalue of the Bloch Hamiltonian $\mathcal{H}_0$.
The equation for $\langle n_T\rangle$ is therefore a familiar linear integral equation and yields
\begin{align}
\langle n_T\rangle^{m}_{\bm{k}}=\tau^m_{\mathrm{tr}\bm{k}} \frac{\nabla T}{T}\cdot \bm{v}^m_{\bm{k}}\left(\varepsilon^m_{\bm{k}}-\mu\right)\frac{\partial f_0(\varepsilon^m_{\bm{k}})}{\partial \varepsilon^m_{\bm{k}}},
\label{rho_T-diagonal}
\end{align}
where $\tau^m_{\mathrm{tr}\bm{k}}$ is the transport lifetime which is often nearly constant across the Fermi surface.  

Next we consider the solution for the off-diagonal part of the density matrix $\langle S_T\rangle$, which is independent of weak disorder.  
From Eq.~(\ref{KE-for-rho_T}) the kinetic equation for $\langle S_T\rangle$ is given by
\begin{align}
\frac{\partial \langle S_T\rangle}{\partial t}+\frac{i}{\hbar}[\mathcal{H}_0,\langle S_T\rangle]=D^{\mathrm{od}}_T(\langle \rho_0\rangle)-J(\langle n_T\rangle),
\end{align}
where $D^{\mathrm{od}}_T(\langle \rho_0\rangle)$ is the off-diagonal part of the intrinsic driving term:
\begin{align}
[D_T(\langle \rho_0\rangle)]^{mm'}_{\bm{k}}=\frac{i}{\hbar}\frac{\nabla T}{T}\cdot\bm{\mathcal{R}}_{\bm{k}}^{mm'}\bigl[\varepsilon^m_{\bm{k}}f_0(\varepsilon^{m}_{\bm{k}})-\varepsilon^{m'}_{\bm{k}}f_0(\varepsilon^{m'}_{\bm{k}})\bigr]
\label{D_T-intrinsic}
\end{align}
with $m\neq m'$.
As one can see from its form, the off-diagonal part of the thermal driving term is responsible for the Berry phase contribution to transport coefficients such as the Nernst conductivity of systems with broken time-reversal symmetry in the absence of a magnetic field.
The solution to this equation is \cite{Culcer2017}
\begin{align}
\langle S_T\rangle=\int_0^\infty dt'\, e^{-i\mathcal{H}_0 t'/\hbar} [D_T(\langle \rho_0\rangle)-J(\langle n_T\rangle)] e^{i\mathcal{H}_0 t'/\hbar},
\end{align}
where we have not explicitly exhibited the time dependences of $\langle \rho_0(t-t')\rangle$ and $\langle n_T(t-t')\rangle$.
It can be further expanded in the eigenstate basis by inserting an infinitesimal $e^{-\eta t'}$ and taking the limit $\eta\rightarrow 0$ to obtain
\begin{align}
\langle S_T\rangle_{\bm{k}}^{mm'}=-i\hbar\frac{[D_T(\langle \rho_0\rangle)]^{mm'}_{\bm{k}} - [J(\langle n_T\rangle)]^{mm'}_{\bm{k}}}{\varepsilon_{\bm{k}}^m-\varepsilon_{\bm{k}}^{m'}}.
\label{rho_T-off-diagonal}
\end{align}
Here, we have written only the principal value part and omitted $\delta$-function terms.
The $\delta$-function terms might become important when bands touch as in the case of electric field \cite{Culcer2017}.
In the case of electric field, such a term gives rise for example to the Zitterbewegung contribution to the minimum conductivity in graphene.
We note that, as shown in Ref.~\cite{Culcer2017}, the contribution from $J(\langle n_T\rangle)$ corresponds to the vertex correction in the ladder-diagram approximation of perturbation theory.

\subsection{General expression for the field-induced density matrix \label{Sec-general-expression-rho}}
Now, we consider the general expression for the density matrix in the presence of electric and magnetic fields and a temperature gradient.
We write the electron density matrix 
as $\langle\rho\rangle=\langle\rho_0\rangle+\langle\rho\rangle_F$, where $\langle\rho_0\rangle$ is the density matrix in the absence of fields, and $\langle\rho\rangle_F$ is the field-induced density matrix.
Then we can rewrite the steady-state uniform limit of Eq.~(\ref{full-kinetic-equation}) at a given wave vector in the form
\begin{align}
(\mathcal{L} - D_E - D_B - D_T)\langle\rho\rangle_F = (D_E+D_B+D_T)\langle \rho_0\rangle,
\end{align}
where we have defined an operator $\mathcal{L}\equiv P + K$.
We have also used the fact that $\mathcal{L}\langle\rho_0\rangle=0$, since $\langle\rho_0\rangle$ in the eigenstate representation is a diagonal matrix and disorder scattering does not occur in the absence of fields.
Here, $P$ acts on an arbitrary density matrix $\langle\rho\rangle$ and is defined by
\begin{align}
P\langle\rho\rangle \equiv \frac{i}{\hbar}[\mathcal{H}_0,\langle \rho\rangle].
\label{definition-of-P}
\end{align}
Note that, in the eigenstate representation, the matrix $P$ is purely diagonal both in wave vector and in density-matrix element at a given wave vector, and that it is nonzero only for off-diagonal density-matrix elements \cite{Sekine2017}.
It follows that
\begin{align}
\langle\rho\rangle_F&=\bigl[1-\mathcal{L}^{-1}(D_E+D_B+D_T)\bigr]^{-1} \mathcal{L}^{-1}(D_E+D_B+D_T)\langle \rho_0\rangle \nonumber\\
&=\sum_{N\ge 0} \bigl[\mathcal{L}^{-1}(D_E+D_B+D_T)\bigr]^N \mathcal{L}^{-1}(D_E+D_B+D_T)\langle \rho_0\rangle.
\label{rho-expansion-general}
\end{align}
We can view the five terms $P$, $K$, $D_E$, $D_B$, and $D_T$ as matrices that act on vectors formed by all 
eigenstate-representation density-matrix components at a given wave vector.
It should be mentioned that Eq.~(\ref{rho-expansion-general}) describes a density-matrix expansion in powers of the field strengths $E_i$, $B_j$, and $-\partial_k T/T$:
\begin{align}
\langle\rho\rangle_F=\sum_{\alpha,\beta,\gamma}\mathcal{M}_{\alpha\beta\gamma}E_i^\alpha B_j^\beta (-\partial_k T/T)^\gamma,
\label{rho-arbitrary-order}
\end{align}
where $i,j,k$ denote the spatial direction, $\alpha,\beta,\gamma$ are integers satisfying $\alpha+\beta+\gamma\ge 1$, and $\mathcal{M}_{\alpha\beta\gamma}$ is a matrix determined from the electronic structure of a system.
Here, note that the angles between the fields are arbitrary in our formalism.
In other words, we can in principle calculate arbitrary-order (linear and nonlinear) responses of a physical observable to the fields from the definition
\begin{align}
\langle\hat{\mathcal{O}}\rangle=\mathrm{Tr}\left[\hat{\mathcal{O}}\langle\rho\rangle_F\right],
\end{align}
where $\hat{\mathcal{O}}$ the operator of a physical observable and $\mathrm{Tr}$ indicates the summation over the wave numbers in the Brillouin zone and over the matrix components.

To be more specific, let us consider the linear response to a temperature gradient $-\nabla T/T$ in the presence of a low magnetic field $\bm{B}$.
From Eq.~(\ref{rho-expansion-general}) we have
\begin{align}
\langle\rho_T\rangle&=\sum_{N,N'\ge 0} (\mathcal{L}^{-1}D_B)^N \mathcal{L}^{-1}D_T (\mathcal{L}^{-1}D_B)^{N'}\langle \rho_0\rangle\nonumber \\
&\equiv\sum_{N\ge 0} (\mathcal{L}^{-1}D_B)^N \mathcal{L}^{-1}D_T(\langle \rho_0\rangle+\langle \rho_B\rangle),
\label{rho-expansion}
\end{align}
Here, the $N=N'=0$ term is given by Eqs.~(\ref{rho_T-diagonal}) and (\ref{rho_T-off-diagonal}), and $\langle \rho_B\rangle$ is the density matrix induced solely by the magnetic field $\langle \rho_B\rangle\equiv\sum_{N\ge 1}(\mathcal{L}^{-1}D_B)^{N}\langle \rho_0\rangle$ \cite{Sekine2017}.
At each order in the magnetic field strength, contributions to $\langle\rho_T\rangle$ can quite generally be organized by their order in an expansion in powers of scattering strength $\lambda$ by letting $K \to \lambda K$ and identifying terms with a particular power of $\lambda$.  
The various low-field expansion terms are generated by repeated action of $D_{B}$ and $\mathcal{L}^{-1}$.
Since we are assuming that the magnetic field $\bm{B}$ is very weak, we may set $\langle \rho_B\rangle=\mathcal{L}^{-1}D_B\langle \rho_0\rangle\equiv\langle\xi_B\rangle$ in Eq.~(\ref{rho-expansion}).
In the eigenstate representation $\langle\xi_B\rangle$ is given by \cite{Sekine2017}
\begin{align}
\langle \xi_B\rangle_{\bm{k}}^{mm}=\frac{e}{\hbar}\, f_0(\varepsilon_{\bm{k}}^m)\, \bm{B}\cdot\bm{\Omega}^m_{\bm{k}},
\label{intrinsic-diagonal-matrix-B}
\end{align}
where $\langle \rho_0\rangle_{\bm{k}}^{mm}=f_0(\varepsilon_{\bm{k}}^m)$ is the Fermi-Dirac distribution function of band $m$, $\Omega_{\bm{k},a}^m=\epsilon^{abc}\, i\langle \partial_{k_b} u_{\bm{k}}^m|\partial_{k_c} u_{\bm{k}}^m\rangle$ is the Berry curvature.
This means that the correction to the Fermi-Dirac distribution function $\langle \rho_0\rangle$ due to magnetic field in Eq.~(\ref{rho-expansion}) is given by the Berry phase correction.

\section{Transport at zero magnetic field \label{Sec-Transport-coefficients}}
At zero magnetic field, an electric current $\bm{J}$ and a heat current $\bm{J}^Q$ in the presence of an electric field $\bm{E}$ and a temperature gradient $-\nabla T/T$ are generally given by
\begin{subequations}
\begin{align}
\bm{J}&=\hat{\sigma}\bm{E}-\hat{\alpha}\nabla T\\
\nonumber\\
\bm{J}^Q&=T\hat{\alpha}\bm{E}-\hat{\kappa}\nabla T,
\end{align}
\label{electric-current-and-heat-current}
\end{subequations}
where $\hat{\sigma}$, $\hat{\alpha}$, and $\hat{\kappa}$ are the electrical conductivity tensor, thermoelectric conductivity tensor, and thermal conductivity tensor, respectively.
The relation between the electric current induced by a temperature gradient ($\bm{J}=-\hat{\alpha}\nabla T$) and the heat current induced by an electric field ($\bm{J}^Q=T\hat{\alpha}\bm{E}$) is known as the Onsager reciprocal relation.

In this section, we derive general expressions for the electric current induced by a temperature gradient and the heat current induced by an electric field at zero magnetic field, focusing on the intrinsic (i.e., Berry phase) contribution to the currents.
We consider a general microscopic model with the Hamiltonian $\mathcal{H}_0=\sum_{m,\bm{k}}(\varepsilon_{\bm{k}}^m-\mu) |m,\bm{k}\rangle\langle m,\bm{k}|$ and the equilibrium density matrix $\langle\rho_0\rangle=\sum_{m,\bm{k}}f_0(\varepsilon_{\bm{k}}^m) |m,\bm{k}\rangle\langle m,\bm{k}|$, where $\varepsilon_{\bm{k}}^m$ is an energy eigenvalue  of band $m$ with momentum $\bm{k}$ and $f_0(\varepsilon_{\bm{k}}^m)=\{\exp[(\varepsilon_{\bm{k}}^m-\mu)/T]+1\}^{-1}$ is the unperturbed Fermi-Dirac distribution function.
In this case, the thermal driving term~(\ref{driving-term-T}) reduces to a little simpler form:
\begin{align}
D_{T}(\langle\rho_0\rangle)=\frac{1}{\hbar}\frac{\nabla T}{T}\cdot\frac{D(\mathcal{H}_0\langle\rho_0\rangle)}{D\bm{k}},
\label{D_T-simpler-form}
\end{align}
since $\mathcal{H}_0$ and $\langle\rho_0\rangle$ are both band-diagonal matrices.

\subsection{Electric current induced by temperature gradient \label{Sec-anomalous-Nernst}}
Let us compute the intrinsic contribution to the electric current induced by a temperature gradient, $\bm{J}=-\hat{\alpha}\nabla T$.
As an example we calculate the intrinsic anomalous Nernst conductivity, which arises as a Berry phase effect.
Without loss of generality, we may consider the case of a temperature gradient along the $y$ direction $-\nabla T/T=-\partial_y T/T\bm{e}_y$, and the current generated along the $x$ direction $J_x$.

First, we compute the following contribution:
\begin{align}
\mathrm{Tr}[(-e)v_x\langle\rho_T\rangle],
\end{align}
where $v_x$ is the velocity in the $x$ direction, and $\langle\rho_T\rangle$ is the density matrix linear in the temperature gradient $-\nabla T/T=-\partial_y T/T\bm{e}_y$.
As we have seen in Sec.~\ref{Sec-rho_T}, the intrinsic contribution that is independent of disorder originates from the off-diagonal component of the thermal driving term.
An off-diagonal component of the thermal driving term~(\ref{D_T-simpler-form}) reads
\begin{align}
&\langle n|D_T(\langle\rho_0\rangle)|n'\rangle\nonumber\\
&=\frac{1}{\hbar}\frac{\partial_y T}{T}\sum_{m'}\varepsilon_{m'}f_{0m'}\langle n|\left[ |\partial_y m'\rangle\langle m'|+|m'\rangle\langle\partial_y  m'|\right]|n'\rangle\nonumber \\
&=\frac{1}{\hbar}\frac{\partial_y T}{T}(\varepsilon_{n'}f_{0n'}-\varepsilon_{n}f_{0n})\langle n|\partial_y n'\rangle,
\end{align}
where $n\neq n'$, $\partial_a\equiv\partial/\partial k_a$, $\varepsilon_m=\varepsilon_{\bm{k}}^m-\mu$, and we have omitted the $\bm{k}$ dependences to simplify the notation.
Using Eq.~(\ref{rho_T-off-diagonal}) we obtain the off-diagonal part of the density matrix induced by the temperature gradient,
\begin{align}
\langle S_T\rangle=-i\frac{\partial_y T}{T}\sum_{nn'}\frac{\varepsilon_{n'}f_{0n'}-\varepsilon_{n}f_{0n}}{\varepsilon_n-\varepsilon_{n'}}|n\rangle\langle n|\partial_y n'\rangle\langle n'|,
\label{S_T-general-expression}
\end{align}
where $n\neq n'$.
We also have the intrinsic contribution to the velocity operator in the eigenstate basis,
\begin{align}
v_x\equiv\frac{1}{\hbar}\frac{D\mathcal{H}_0}{Dk_x}=\frac{1}{\hbar}\sum_{m'}(\varepsilon_{m'}-\varepsilon_{n'})\left[|\partial_x m'\rangle\langle m'|+|m'\rangle\langle \partial_x m'|\right],
\label{v_x-general-expression}
\end{align}
where we have used the fact that $\varepsilon_{n'}\partial_x(\sum_{m'}|m'\rangle\langle m'|)=0$.
Note that the terms proportional to $\partial_x\varepsilon$ in $D\mathcal{H}_0/Dk_x$ do not contribute to the final expression for the current resulting from $\langle S_T\rangle$ due to the traceless nature.
From Eqs.~(\ref{S_T-general-expression}) and (\ref{v_x-general-expression}) we obtain
\begin{align}
&\langle m|v_x\langle S_T\rangle|m\rangle\nonumber\\
&=-\frac{i}{\hbar}\frac{\partial_y T}{T}\sum_{nn'}\sum_{m'}(\varepsilon_{m'}-\varepsilon_{n'})\frac{\varepsilon_{n'}f_{0n'}-\varepsilon_{n}f_{0n}}{\varepsilon_n-\varepsilon_{n'}}\delta_{n'm}\nonumber\\
&\ \ \ \ \  \times\left[\delta_{m'n}\langle m|\partial_x n\rangle\langle n|\partial_y m\rangle+\delta_{mm'}\langle \partial_x m|n\rangle\langle n|\partial_y m\rangle\right] \nonumber\\
&=\frac{i}{\hbar}\frac{\partial_y T}{T}\left[\varepsilon_{m}f_{0m}\langle\partial_x m|\partial_y m\rangle+\sum_n\varepsilon_{n}f_{0n}\langle m|\partial_x n\rangle\langle n|\partial_y m\rangle\right].
\end{align}
Finally, we obtain the electronic contribution to the electric current induced by a temperature gradient,
\begin{align}
\mathrm{Tr}[(-e)v_x\langle S_T\rangle]=-\frac{e}{\hbar}\frac{\partial_y T}{T}\sum_m\int[d\bm{k}]\Omega_{\bm{k},z}^m(\varepsilon_{\bm{k}}^m-\mu)f_0(\varepsilon_{\bm{k}}^m),
\label{electric-current1}
\end{align}
where $\int[d\bm{k}]=\int d^dk/(2\pi)^d$ and $\Omega^m_a=\epsilon_{abc}i\langle\partial_{k_a} m|\partial_{k_b} m\rangle$ is the Berry curvature of band $m$.
Note that the contribution from the diagonal density matrix $\langle n_T\rangle$ to Eq.~(\ref{electric-current1}) is zero, because the integrand is an odd function of the wave vector $\bm{k}$.

Next, we need to calculate the contribution from the magnetization, $-\bm{E}_T\times\bm{M}$, which is analogous to that in case of electric field, $-\bm{E}\times\bm{M}$.
We calculate the intrinsic magnetization (i.e., orbital magnetization) $\bm{M}$ using the identity $\bm{M}=-\partial F/\partial \bm{B}|_{\bm{B}\to 0}$ with $F$ and $\bm{B}$ being the grand canonical potential of the system and a magnetic field, respectively.
In our quantum kinetic formalism, it is the equilibrium electron density matrix that is modified by a magnetic field in systems with momentum-space Berry curvatures [see Eq,~(\ref{intrinsic-diagonal-matrix-B})], while the momentum-space density of states remains unchanged \cite{Sekine2017}.
Then, the total number of electrons of the system in a magnetic field reads \cite{Sekine2017}
\begin{align}
N=\mathrm{Tr}[(\langle \rho_0\rangle+\langle \xi_B\rangle)]=\sum_m\int[d\bm{k}]\left(1+\frac{e}{\hbar}\bm{B}\cdot\bm{\Omega}_{\bm{k}}^m\right)f_0(\varepsilon_{\bm{k}}^m),
\end{align}
where $\langle \xi_B\rangle=(e/\hbar)f_0(\varepsilon_{\bm{k}}^m)\, \bm{B}\cdot\bm{\Omega}^m_{\bm{k}}$ [Eq.~(\ref{intrinsic-diagonal-matrix-B})].
The thermodynamic relation $N=-\partial F/\partial \mu$ with $\mu$ being the chemical potential can be rewritten as $F=-\int d\mu\, N$.
Then, we obtain
\begin{align}
F&=-\int d\mu\sum_m\int[d\bm{k}]\left(1+\frac{e}{\hbar}\bm{B}\cdot\bm{\Omega}_{\bm{k}}^m\right)f_0(\varepsilon_{\bm{k}}^m)\nonumber \\
&=-\frac{1}{\beta}\sum_m\int[d\bm{k}]\left(1+\frac{e}{\hbar}\bm{B}\cdot\bm{\Omega}_{\bm{k}}^m\right)\ln \left[1+e^{-\beta(\varepsilon_{\bm{k}}^m-\mu)}\right],
\label{free-energy}
\end{align}
where we have used that $\int d\mu f_0(\varepsilon_{\bm{k}}^m)=(1/\beta)\ln[1+e^{-\beta(\varepsilon_{\bm{k}}^m-\mu)}]$ with $\beta=1/T$, because the well-known thermodynamic identity $F=(1/\beta)\sum_m\int[d\bm{k}]\ln[1+e^{-\beta(\varepsilon_{\bm{k}}^m-\mu)}]$ holds in the limit $\bm{B}\to 0$.
Finally, we get
\begin{align}
\bm{M}=\frac{e}{\hbar}\frac{1}{\beta}\sum_m\int[d\bm{k}]\bm{\Omega}_{\bm{k}}^m\ln \left[1+e^{-\beta(\varepsilon_{\bm{k}}^m-\mu)}\right],
\label{intrinsic-magnetization}
\end{align}
which is in complete agreement with the expression obtained by invoking semiclassical wave-packet dynamics \cite{Xiao2006}.

In order to obtain the ``transport'' current, the contribution from the orbital magnetization must be subtracted, since it flows even in equilibrium.
Finally, from Eqs.~(\ref{electric-current1}) and (\ref{intrinsic-magnetization}), the ``transport'' electric current is obtained as
\begin{align}
J_x=&\ \mathrm{Tr}[(-e)v_x\langle S_T\rangle]-(-\bm{E}_T\times\bm{M})\nonumber\\
=&-\frac{e}{\hbar}\frac{\partial_y T}{T}\sum_m\int[d\bm{k}]\Omega_{\bm{k},z}^m\nonumber\\
&\times\left\{(\varepsilon_{\bm{k}}^m-\mu)f_0(\varepsilon_{\bm{k}}^m)+T\ln \left[1+e^{-\beta(\varepsilon_{\bm{k}}^m-\mu)}\right]\right\},
\label{electric-current-general-expression}
\end{align}
which is in complete agreement with the expression obtained by invoking semiclassical wave-packet dynamics \cite{Xiao2006}.
Note that we have not included the correction to the Bloch-state energy due to the orbital magnetic moment $\bm{\mathfrak{m}}_{\bm{k}}^m$ [i.e., the modification such that $\varepsilon_{\bm{k}}^m \to\varepsilon_{\bm{k}}^m-\bm{\mathfrak{m}}_{\bm{k}}^m\cdot\bm{B}$ in Eq.~(\ref{free-energy})], which adds the contribution $\sum_m\int[d\bm{k}]f_0(\varepsilon_{\bm{k}}^m)\bm{\mathfrak{m}}_{\bm{k}}^m$ to the total orbital magnetization in Eq.~(\ref{intrinsic-magnetization}).
This is because the contribution from the orbital magnetic moment does not appear explicitly in the final expression for the ``transport'' current at zero magnetic field.
In other words, this contribution is cancelled out by the ``local'' current $-\bm{E}_T\times\sum_m\int[d\bm{k}]f_0(\varepsilon_{\bm{k}}^m)\bm{\mathfrak{m}}_{\bm{k}}^m$ which should be added to the right-hand side of Eq.~(\ref{electric-current-general-expression}) [see Eq.~(11) of Ref.~\cite{Xiao2006}].

\subsection{Heat current induced by electric field}
\begingroup
\renewcommand{\arraystretch}{1.6}
\begin{table*}[!t]
\caption{Schematic comparison of our formalism with semiclassical wave-packet dynamics in the presence of a temperature gradient $\bm{E}_T=-\nabla T/T$ and an electric field $\bm{E}$.
The orbital magnetization $\bm{M}$ is given by Eq.~(\ref{intrinsic-magnetization}).
Note that in this table the contributions from the orbital magnetic moment $\bm{\mathfrak{m}}$ are already canceled out and not shown in the expressions for the currents obtained by invoking semiclassical wave-packet dynamics.
}
\begin{ruledtabular}
\begin{tabular}{ccc}
Formalism & Electric current ($\bm{J}=-\hat{\alpha}\nabla T$) & Heat current ($\bm{J}^Q=T\hat{\alpha}\bm{E}$)\\
\hline
Semiclassical wave-packet dynamics \cite{Xiao2006} & $\nabla\times\bm{M}$ & $\int[d\bm{k}](\varepsilon-\mu)\dot{\bm{r}}f_0+\bm{E}\times\bm{M}$\\
Our quantum kinetic formalism & $\mathrm{Tr}[(-e)\bm{v}\langle S_T\rangle]+\bm{E}_T\times\bm{M}$ & $\mathrm{Tr}\left[\tfrac{1}{2}\{\mathcal{H}_0, \bm{v}\}\langle S_E\rangle\right]-\mathrm{Tr}\left[(\bm{E}\times\bm{\mathfrak{m}})\langle \rho_0\rangle\right]+\bm{E}\times\bm{M}$
\end{tabular}
\end{ruledtabular}\label{Table1}
\end{table*}
\endgroup
Let us compute the intrinsic contribution to the heat current induced by an electric field, $\bm{J}^Q=T\hat{\alpha}\bm{E}$.
To see whether our theory correctly describes the Onsager reciprocal relation in Eq.~(\ref{electric-current-and-heat-current}), we consider the reciprocal effect of the anomalous Nernst effect, i.e., a heat current generation along the $x$ direction $J^Q_x$ by an electric field along the $y$ direction $\bm{E}=E_y\bm{e}_y$.

First, we compute the following contribution:
\begin{align}
\mathrm{Tr}\left[\tfrac{1}{2}\{\mathcal{H}_0, v_x\}\langle\rho_E\rangle\right],
\end{align}
where $\frac{1}{2}\{\mathcal{H}_0, v_x\}$ is the energy current operator in the $x$ direction, and $\langle\rho_E\rangle$ is the density matrix linear in the electric field $\bm{E}=E_y\bm{e}_y$.
By comparing the thermal driving term~(\ref{driving-term-T}) [or Eq.~(\ref{D_T-simpler-form}) in the present case] and the electric driving term~(\ref{driving-term-E}), we see that these driving terms have a very similar structure.
Then, from Eq.~(\ref{S_T-general-expression}) we find that the off-diagonal part of the density matrix induced by the electric field is given by
\begin{align}
\langle S_E\rangle=-ieE_y\sum_{n,n'}\frac{f_{0n'}-f_{0n}}{\varepsilon_n-\varepsilon_{n'}}|n\rangle\langle n|\partial_y n'\rangle\langle n'|,
\label{S_E-general-expression}
\end{align}
where $n\neq n'$.
From Eqs.~(\ref{v_x-general-expression}) and (\ref{S_E-general-expression}) we obtain
\begin{align}
\langle m|\mathcal{H}_0v_x\langle S_E\rangle|m\rangle
=i\frac{eE_y}{\hbar}\sum_n\varepsilon_{m}\left(f_{0n}-f_{0m}\right)\langle m|\partial_x n\rangle\langle n|\partial_y m\rangle.
\end{align}
Similarly, we can calculate $\langle m|v_x\mathcal{H}_0\langle S_E\rangle|m\rangle$.
Then, combining these things, we have the following general expression:
\begin{align}
&\mathrm{Tr}\left[\tfrac{1}{2}\{\mathcal{H}_0, v_x\}\langle S_E\rangle\right]\nonumber\\
&=i\frac{eE_y}{2\hbar}\sum_{m,n}(\varepsilon_{m}+\varepsilon_{n})\left(f_{0n}-f_{0m}\right)\langle m|\partial_x n\rangle\langle n|\partial_y m\rangle\nonumber\\
&=\frac{eE_y}{2\hbar}\left[\sum_{m}\varepsilon_{m}\Omega_z^m f_{0m}+i\sum_{m,n}\left(\varepsilon_{m}f_{0n}-\varepsilon_{n}f_{0m}\right)\langle m|\partial_x n\rangle\langle n|\partial_y m\rangle\right].
\label{H_0-v_x-S_E}
\end{align} 
Here, let us introduce the orbital magnetic moment of an electron in band $n$ defined by \cite{Xiao2010}
\begin{align}
\bm{\mathfrak{m}}_{\bm{k}}^n&=-i\frac{e}{2\hbar}\langle\nabla_{\bm{k}} n|\times[\mathcal{H}_0-\varepsilon_n]|\nabla_{\bm{k}}n\rangle\nonumber\\
&=i\frac{e}{2\hbar}\sum_{m'}(\varepsilon_{m'}-\varepsilon_{n})\langle n|\nabla_{\bm{k}}m'\rangle\times\langle m'|\nabla_{\bm{k}}n\rangle,
\label{orbital-magnetic-moment}
\end{align}
where we have used that $\nabla_{\bm{k}}(\langle n|m'\rangle)=0$.
Using Eq.~(\ref{orbital-magnetic-moment}) and the identities $\varepsilon_{m}=(\varepsilon_{m}-\varepsilon_{n})+\varepsilon_{n}$ and $\varepsilon_{n}=(\varepsilon_{n}-\varepsilon_{m})+\varepsilon_{m}$ in the second term in the right-hand side of Eq.~(\ref{H_0-v_x-S_E}), we can rewrite Eq.~(\ref{H_0-v_x-S_E}) as
\begin{align}
\mathrm{Tr}\left[\tfrac{1}{2}\{\mathcal{H}_0, v_x\}\langle S_E\rangle\right]
=&\ \frac{eE_y}{\hbar}\sum_{m}\int[d\bm{k}]\Omega_{\bm{k},z}^m(\varepsilon_{\bm{k}}^m-\mu)f_0(\varepsilon_{\bm{k}}^m)\nonumber\\
&+E_y\sum_n\int[d\bm{k}]\mathfrak{m}_{\bm{k},z}^n f_0(\varepsilon_{\bm{k}}^n),
\label{heat-current-from-S_E}
\end{align}
which indicates that the contribution from the orbital magnetic moment, $-\mathrm{Tr}[(\bm{E}\times\bm{\mathfrak{m}})_x\langle \rho_0\rangle]$, is {\it already subtracted} in the definition of $\mathrm{Tr}[\frac{1}{2}\{\mathcal{H}_0, \bm{v}\}\langle S_E\rangle]$.
Then, it turns out that the electronic part of the energy current in our formalism should be defined as $\mathrm{Tr}[\frac{1}{2}\{\mathcal{H}_0, \bm{v}\}\langle S_E\rangle]-\mathrm{Tr}[(\bm{E}\times\bm{\mathfrak{m}})\langle \rho_0\rangle]$.
This form corresponds to the energy current carried by a wave packet, $\langle W|\frac{1}{2}\{\hat{H}, \hat{\dot{\bm{r}}}\}|W\rangle=\varepsilon\dot{\bm{r}}-\bm{E}\times\bm{\mathfrak{m}}$, in the language of semiclassical wave-packet dynamics \cite{Xiao2006}.

In order to obtain the ``transport'' current, the contribution from the orbital magnetization, $-\bm{E}\times\bm{M}$ (i.e., an energy flow due to the Poynting vector), must be subtracted, since it flows even in equilibrium \cite{Xiao2006}.
The orbital magnetization $\bm{M}$ has already been calculated in Eq.~(\ref{intrinsic-magnetization}).
Finally, we obtain the ``transport'' heat current as
\begin{align}
J^Q_x=&\ \mathrm{Tr}\left[\tfrac{1}{2}\{\mathcal{H}_0, v_x\}\langle S_E\rangle\right]-\mathrm{Tr}\left[(\bm{E}\times\bm{\mathfrak{m}})_x\langle \rho_0\rangle\right]+\bm{E}\times\bm{M}\nonumber\\
=&\ \frac{eE_y}{\hbar}\sum_m\int[d\bm{k}]\Omega_{\bm{k},z}^m\nonumber\\
&\ \times\left\{(\varepsilon_{\bm{k}}^m-\mu)f_0(\varepsilon_{\bm{k}}^m)+T\ln \left[1+e^{-\beta(\varepsilon_{\bm{k}}^m-\mu)}\right]\right\}.
\label{heat-current-general-expression}
\end{align}
Comparing Eqs.~(\ref{electric-current-general-expression}) and (\ref{heat-current-general-expression}), we confirm the Onsager reciprocal relation $\alpha_{xy}=J_x/(-\partial_y T)=J^Q_x/(TE_y)$, as expected in Eq.~(\ref{electric-current-and-heat-current}).
We show a schematic comparison of our formalism with semiclassical wave-packet dynamics in the presence of a temperature gradient $-\nabla T/T$ and an electric field $\bm{E}$ in Table~\ref{Table1}.

\subsection{Heat current induced by temperature gradient}
Let us consider the intrinsic contribution to the heat current induced by a temperature gradient, $\bm{J}^Q=-\hat{\kappa}\nabla T$.
As an example, we consider the anomalous thermal Hall effect.
Without loss of generality, we may consider the case of a temperature gradient along the $y$ direction $-\nabla T/T=-\partial_y T/T\bm{e}_y$, and the heat current generated along the $x$ direction $J_x^Q$.

First, there is the contribution from the usual energy current operator $\hat{j}_x^Q=\frac{1}{2}\{\mathcal{H}_0,v_x\}$, which is given by $\mathrm{Tr}[\frac{1}{2}\{\mathcal{H}_0,v_x\}\langle S_T\rangle]$, where $\langle S_T\rangle$ is the intrinsic density matrix linear in the temperature gradient $-\nabla T/T=-\partial_y T/T\bm{e}_y$.
Comparing $\langle S_T\rangle$ [Eq.~(\ref{S_T-general-expression})] and $\langle S_E\rangle$ [Eq.~(\ref{S_E-general-expression})], we can see that a replacement such that $E_y\to \partial_y T/T$ and $f_{0n}\to \varepsilon_nf_{0n}$ in $\langle S_E\rangle$ gives rise to $\langle S_T\rangle$.
Then, from Eq.~(\ref{heat-current-from-S_E}) we find that
\begin{align}
\mathrm{Tr}\left[\tfrac{1}{2}\{\mathcal{H}_0, v_x\}\langle S_T\rangle\right]
=&\ \frac{e}{\hbar}\frac{\partial_y T}{T}\sum_{m}\int[d\bm{k}]\Omega_{\bm{k},z}^m(\varepsilon_{\bm{k}}^m-\mu)^2f_0(\varepsilon_{\bm{k}}^m)\nonumber\\
&+\frac{\partial_y T}{T}\sum_n\int[d\bm{k}]\mathfrak{m}_{\bm{k},z}^n (\varepsilon_{\bm{k}}^n-\mu)f_0(\varepsilon_{\bm{k}}^n),
\label{heat-current-S_T}
\end{align}
where it is indicated that the energy flow due to the orbital magnetic moment has already been subtracted.
As in the case of the heat current induced by an electric field [Eq.~(\ref{heat-current-general-expression})], the contribution from the orbital magnetic moment should not appear in the transport current.
Thus, the electronic part of the energy current in our formalism should be defined as $-\mathrm{Tr}[\frac{1}{2}\{\mathcal{H}_0, \bm{v}\}\langle S_T\rangle]-\mathrm{Tr}[(\bm{E}_T\times\bm{\mathfrak{m}})_x\mathcal{H}_0\langle\rho_0\rangle]$.

In addition to Eq.~(\ref{heat-current-S_T}), it has been shown that there is the contribution from the energy magnetization $\bm{M}_E$ to the anomalous thermal Hall effect \cite{Qin2011,Sumiyoshi2013,Shitade2014,Gromov2015,Nakai2016}.
From the calculation of the anomalous Nernst effect in Sec.~\ref{Sec-anomalous-Nernst}, we expect that such a contribution from the energy magnetization should take the form $-\bm{E}_T\times\bm{M}_E$.
Finally, the anomalous thermal Hall effect in our formalism is expected to be given by
\begin{align}
J^Q_x=&-\mathrm{Tr}\left[\tfrac{1}{2}\{\mathcal{H}_0, v_x\}\langle S_T\rangle\right]-\mathrm{Tr}\left[(\bm{E}_T\times\bm{\mathfrak{m}})_x\mathcal{H}_0\langle \rho_0\rangle\right]\nonumber\\
&-(-\bm{E}_T\times\bm{M}_E).
\end{align}
However, it has been suggested that the calculation of the energy magnetization $\bm{M}_E$ is complicated \cite{Qin2011,Sumiyoshi2013,Shitade2014,Gromov2015,Nakai2016}, and is beyond the scope of this paper since we are focusing on the thermoelectric and thermal trasnport in a magnetic field.

\section{Thermal chiral anomaly in generic three-dimensional semimetals \label{Sec-CA-generic}}
So far we have considered intrinsic transport in systems with momentum-space Berry curvatures in the absence of a magnetic field.
Now, we take into account the presence of a magnetic field in transport phenomena induced by a temperature gradient.
We take the chiral anomaly in Weyl semimetals as a representative example.
The chiral anomaly in Weyl semimetals is usually referred to as the non-conservation of the total number of electrons in a given valley (Weyl cone) in the presence of parallel electric and magnetic fields.
In semiclassical wave-packet dynamics, it has been shown that such a non-conservation of the total number of electrons in a given valley also happens in the presence of parallel temperature gradient and magnetic field \cite{Spivak2016}.
Namely, the thermal chiral anomaly can occur in 3D semimetals.
In this section, we explicitly calculate the rate of the change of the total electron number in a given valley (the rate of pumping of electrons between valley) in a generic model of 3D semimetals.
We set $\hbar= 1$ in the rest of this paper.

We study a general model with the Hamiltonian $\mathcal{H}_0=\sum_{m,\bm{k}}\varepsilon_{\bm{k}}^m |m,\bm{k}\rangle\langle m,\bm{k}|$ and the equilibrium density matrix $\langle\rho_0\rangle=\sum_{m,\bm{k}}f_0(\varepsilon_{\bm{k}}^m) |m,\bm{k}\rangle\langle m,\bm{k}|$, where $\varepsilon_{\bm{k}}^m$ is an energy eigenvalue of band $m$ with momentum $\bm{k}$ and $f_0(\varepsilon_{\bm{k}}^m)$ is the Fermi-Dirac distribution function.
For concreteness and without loss of generality, we may choose $\bm{E}_T=(0,0,-\partial_z T/T)$ and $\bm{B}=(0,0,B_z)$.
Let us consider the following quantity that is linear in both temperature gradient and magnetic field,
\begin{align}
\frac{\partial N}{\partial t}\equiv\mathrm{Tr}[\mathcal{L}\langle\rho_{TB}\rangle]=\mathrm{Tr}\left[D_B\mathcal{L}^{-1} D_T(\langle \rho_0\rangle) + D_T\mathcal{L}^{-1} D_B(\langle \rho_0\rangle)\right]
\label{Rate-of-pumping}
\end{align}
evaluated for the Fermi surface associated with a particular valley.
Here, the density matrix $\langle\rho_{TB}\rangle$ is obtained from Eq.~(\ref{rho-expansion}).
Note that $\mathcal{L}=P+K$ is an operator introduced in Sec.~\ref{Sec-general-expression-rho}, which has the dimension of [time]$^{-1}$.

Let us consider the first term in the right-hand side of Eq.~(\ref{Rate-of-pumping}), i.e., $\mathrm{Tr}[D_B\mathcal{L}^{-1} D_T(\langle \rho_0\rangle)]$.
The density matrix linear in the temperature gradient, $\langle\rho_T\rangle=\mathcal{L}^{-1} D_T(\langle \rho_0\rangle)$, contains both band-diagonal $\langle n_T\rangle$ and band off-diagonal $\langle S_T\rangle$ contributions as $\langle\rho_T\rangle=\langle n_T\rangle+\langle S_T\rangle$.
Then, it follows that $\mathrm{Tr}[D_B \mathcal{L}^{-1} D_T(\langle \rho_0\rangle)]=\mathrm{Tr}[D_B(\langle n_T\rangle)]+\mathrm{Tr}[D_B(\langle S_T\rangle)]$, since $D_B$ is linear in the density matrix.
First we evaluate the diagonal element $[D_B(\langle n_T\rangle)]^{mm}_{\bm{k}}$.
From Eq.~(\ref{rho_T-diagonal}) we have $\langle n_T\rangle=-E_{T,z}\sum_{m,\bm{k}}\tau_{\mathrm{tr}}^m[\partial_{k_z} f_0(\varepsilon_{\bm{k}}^m)] |m,\bm{k}\rangle\langle m,\bm{k}|$.
It can be shown that for arbitrary diagonal density matrix $\langle \mathcal{F}\rangle$ the diagonal component of the magnetic driving term $[D_B(\langle \mathcal{F}\rangle)]$ is (see Appendix~\ref{Appendix-D_B-1})
\begin{align}
[D_B(\langle \mathcal{F}\rangle)]^{mm}_{\bm{k}}=eB_z\left(\frac{\partial \varepsilon_{\bm{k}}^m}{\partial k_y}\frac{\partial}{\partial k_x}-\frac{\partial \varepsilon_{\bm{k}}^m}{\partial k_x}\frac{\partial}{\partial k_y}\right)\mathcal{F}_{\bm{k}}^m,
\end{align}
where $\langle \mathcal{F}\rangle^{mm}_{\bm{k}}=\mathcal{F}_{\bm{k}}^m$.
In the present case we can set $\mathcal{F}_{\bm{k}}^m=-E_{T,z}\tau_{\mathrm{tr}}^m\partial_{k_z} f_0(\varepsilon_{\bm{k}}^m)$.
Then, it is obvious that $[D_B(\langle n_T\rangle)]^{mm}_{\bm{k}}$ is an odd function of $k_x$, $k_y$, and $k_z$, which means that $\mathrm{Tr}\, [D_B(\langle n_T\rangle)]=0$.
Therefore, it turns out that one of the two contributions to the rate of the change of the total electron number in a given valley due to the thermal chiral anomaly [Eq.~(\ref{Rate-of-pumping})] is given by (see Appendix~\ref{dN/dt-analytical})
\begin{align}
&\mathrm{Tr}\left[D_B\mathcal{L}^{-1} D_T(\langle \rho_0\rangle)\right]\nonumber\\
&=\int_{\mathrm{FS}} \frac{d^3k}{(2\pi)^3}\, \sum_m\left[D_B(\langle  S_T\rangle)\right]^{mm}_{\bm{k}}\nonumber\\
&=\frac{eB_z\partial_zT}{4\pi^2}\int \frac{d^3 k}{2 \pi}\frac{\varepsilon^m_{\bm{k}}-\mu}{T}\frac{\partial f_{0}(\varepsilon_{\bm{k}}^m)}{\partial \varepsilon^m_{\bm{k}}}\big[ v^m_{\bm{k},x} \Omega^m_{\bm{k},x} + v^m_{\bm{k},y} \Omega^m_{\bm{k},y} \big],
\label{dN/dt-1}
\end{align}
where FS represents the integration on the Fermi surface of the valley, $\Omega_{\bm{k},a}^m=\epsilon^{abc}\, i\langle \partial_{k_b} u_{\bm{k}}^m|\partial_{k_c} u_{\bm{k}}^m\rangle$ is the Berry curvature of band $m$, and $\bm{v}^m_{\bm{k}}$ is the Bloch state group velocity.
In Eq.~(\ref{dN/dt-1}) we have assumed that only the band $m$ intersects the Fermi surface, i.e., $\partial f_0(\varepsilon^n_{\bm{k}})/\partial \varepsilon^n_{\bm{k}}=\delta_{mn}\partial f_0(\varepsilon^m_{\bm{k}})/\partial \varepsilon^m_{\bm{k}}$, which can in general apply to multi-valley systems.

Next let us consider the second term in the right-hand side of Eq.~(\ref{Rate-of-pumping}), i.e., $\mathrm{Tr}[D_T\mathcal{L}^{-1} D_B(\langle \rho_0\rangle)]$.
As has been explained in Sec.~\ref{Sec-general-expression-rho}, the linear-response density matrix to a low magnetic field in the absence of a temperature gradient, $\langle\rho_B\rangle=\mathcal{L}^{-1} D_{B}(\langle \rho_0\rangle)$,
contains only band-diagonal contribution as $\langle\rho_B\rangle=\langle \xi_B\rangle$.
In this case the calculation is much easier than that of $\mathrm{Tr}[D_B(\langle\rho_T\rangle)]$.
After a calculation we find that
\begin{align}
\mathrm{Tr}\left[D_T\mathcal{L}^{-1} D_B(\langle \rho_0\rangle)\right]
&=\int_{\mathrm{FS}} \frac{d^3k}{(2\pi)^3}\, \sum_m\left[D_T(\langle\xi_B\rangle)\right]^{mm}_{\bm{k}} \nonumber\\
&=\frac{eB_z\partial_zT}{4\pi^2}\int \frac{d^3 k}{2 \pi}\frac{\varepsilon^m_{\bm{k}}-\mu}{T}\frac{\partial f_{0}(\varepsilon_{\bm{k}}^m)}{\partial \varepsilon^m_{\bm{k}}}v^m_{\bm{k},z} \Omega^m_{\bm{k},z},
\label{dN/dt-2}
\end{align}
where we have assumed again that only the band $m$ intersects the Fermi surface.

Combining Eqs.~(\ref{dN/dt-1}) and (\ref{dN/dt-2}) we arrive at the final expression for the rate of pumping of electrons between valleys due to the thermal chiral anomaly:
\begin{align}
\frac{\partial N}{\partial t}
&=\mathrm{Tr}\left[D_B(\langle S_T\rangle)\right]+\mathrm{Tr}\left[D_T(\langle \xi_B\rangle)\right] \nonumber\\
&=\frac{eB_z\partial_z T}{4\pi^2}\int \frac{d^3 k}{2 \pi}\frac{\varepsilon^m_{\bm{k}}-\mu}{T}\frac{\partial f_{0}(\varepsilon_{\bm{k}}^m)}{\partial \varepsilon^m_{\bm{k}}}\bm{v}^m_{\bm{k}} \cdot \bm{\Omega}^m_{\bm{k}},
\label{dN/dt-total}
\end{align}
which can be regarded as the generalization of the expression obtained by disorder-free semiclassical wave-packet dynamics \cite{Spivak2016}.
To the best of our knowledge, a microscopic derivation of the rate of pumping, Eq.~(\ref{dN/dt-total}), has not yet been done in a general quantum model.
We see that Eq.~(\ref{dN/dt-total}) resembles the expression for the rate of pumping in parallel electric and magnetic fields due to the chiral anomaly \cite{Sekine2017}:
\begin{align}
\frac{\partial N}{\partial t}
=\frac{e^2 E_zB_z}{4\pi^2}\int \frac{d^3 k}{2 \pi}\frac{\partial f_{0}(\varepsilon_{\bm{k}}^m)}{\partial \varepsilon^m_{\bm{k}}}\bm{v}^m_{\bm{k}} \cdot \bm{\Omega}^m_{\bm{k}}.
\label{dN/dt-EB}
\end{align}
However, a big difference is the presence of the factor $(\varepsilon^m_{\bm{k}}-\mu)/T$ due to which Eq.~(\ref{dN/dt-total}) approaches zero in the zero temperature limit $T/\mu\to 0$, while Eq.~(\ref{dN/dt-EB}) can be finite even at zero temperature.

Let us take a closer look at the relation between the rate of pumping induced by a temperature gradient [Eq.~(\ref{dN/dt-total})] and that by an electric field [Eq.~(\ref{dN/dt-EB})].
To this end, we rewrite Eqs.~(\ref{dN/dt-total}) and (\ref{dN/dt-EB}) in a unified fashion as
\begin{align}
\frac{\partial N}{\partial t}
=\mathcal{N}_{EB}E_z-\mathcal{N}_{TB}\partial_zT,
\end{align}
which can be viewed as an analogy to an electric current in the presence of an electric field and a temperature gradient, $\bm{J}=\hat{\sigma}\bm{E}-\hat{\alpha}\nabla T$.
Using the Sommerfeld expansion, we have at low temperatures $T\ll \mu$
\begin{align}
\mathcal{N}_{TB}
&=-\frac{eB_z}{4\pi^2}\int d\varepsilon\, Q(\varepsilon)\frac{\varepsilon-\mu}{T}\frac{\partial f_{0}(\varepsilon)}{\partial \varepsilon}\nonumber\\
&=\frac{eB_z}{4\pi^2}\frac{\pi^2}{3}T \left.\frac{\partial Q(\mu)}{\partial \mu}\right|_{T=0}\nonumber\\
&=\frac{\pi^2}{3e}T\left.\frac{\partial\mathcal{N}_{EB}}{\partial \mu}\right|_{T=0},
\label{dN/dt-Mott-relation}
\end{align}
where $Q(\mu)=\int d^3 k/(2 \pi)\, \delta(\mu-\varepsilon_{\bm{k}}^m)\bm{v}^m_{\bm{k}} \cdot \bm{\Omega}^m_{\bm{k}}$.
From Eq.~(\ref{dN/dt-Mott-relation}) we find that the ``Mott relation'' between $\mathcal{N}_{TB}$ and $\mathcal{N}_{EB}$ is satisfied.
Note, however, that the rate of pumping itself is not a physical observable.
Namely, the integrand in Eq.~(\ref{dN/dt-total}) appears at intermediate steps of calculation processes of physical observables such as thermoelectric and thermal conductivities, as we shall see in Sec.~\ref{WSM-thermoelectric}.

In closing, it is informative to consider a simple 3D Weyl Hamiltonian, $\mathcal{H}(\bm{k})=v_F(k_x\sigma_x+k_y\sigma_y+k_z\sigma_z)$, for which we obtain the energy eigenvalues $\varepsilon^\pm_{\bm{k}}=\pm\varepsilon_{\bm{k}}=\pm v_F\sqrt{k_x^2+k_y^2+k_z^2}$ and the Berry curvature $\Omega^\pm_{\bm{k},a}=\mp v_F^3 k_a/2\varepsilon^3_{\bm{k}}$ ($a=x,y,z$) (see Sec.~\ref{Theoretical-model} for details).
In this model, $Q(\mu)$ is independent of $\mu$ for each valley (Weyl cone), and therefore $\partial Q(\mu)/\partial \mu=0$.
This means that $\mathcal{N}_{TB}=0$ at the lowest order in the Sommerfeld expansion.
Actually, $\mathcal{N}_{TB}$ behaves as $\sim e^{-\mu/T}$ and rapidly approaches zero at low temperatures $T\ll \mu$.
However, $\partial Q(\mu)/\partial \mu$ can be nonzero in more realistic models whose dispersions have the higher-order terms in $k_a$.

\section{Application to Weyl semimetals \label{Sec-WSM}}
In this section, as an application of our theory, we study longitudinal thermoelectric and thermal transport in Weyl semimetals in a magnetic field.
In order to investigate the relations of thermoelectric and heat currents to electric current (i.e., the Mott relation and the Wiedemann-Franz law, respectively), we focus on the longitudinal thermoelectric and thermal conductivities quadratic in magnetic field which are induced by the thermal chiral anomaly.
Namely, we investigate the relations of thermoelectric and heat currents to the positive quadratic magnetoconductivity arising from the chiral anomaly \cite{Son2013,Burkov2014,Sekine2017}
\begin{align}
\sigma_{zz}^{\mathrm{CA}}=\frac{e^2}{8\pi^2}\frac{(eB_z)^2v_F^3}{\mu^2}\tau,
\label{Sigma_zz-Son&Spivak}
\end{align}
where $v_F$ is the Fermi velocity, $\mu$ is the chiral potential, and $\tau$ is the intervalley scattering time.
As we have seen in Sec.~\ref{Sec-Transport-coefficients}, the magnetic-field dependent contributions to the electric and heat currents are respectively calculated from
\begin{subequations}
\begin{align}
J_i&=\mathrm{Tr}[(-e)v_i\langle\rho_{TB^2}\rangle],\\
\nonumber\\
J^Q_i&=\mathrm{Tr}\left[\tfrac{1}{2}\{\mathcal{H}_0, v_i\}\langle\rho_{TB^2}\rangle\right],
\end{align}
\label{Electric-and-heat-current}
\end{subequations}
where $\langle\rho_{TB^2}\rangle$ is the density matrix that is linear in temperature gradient and quadratic in magnetic field.
Here, note that the contributions from the orbital magnetic moment $\bm{\mathfrak{m}}$ (i.e., $\mathrm{Tr}[(\bm{E}\times\bm{\mathfrak{m}})\langle \rho_0\rangle]$) and the orbital magnetization $\bm{M}$ (i.e., $\bm{E}_T\times\bm{M}$ and $\bm{E}\times\bm{M}$)
to the electric and heat currents are present only in the transverse currents, as is readily understood from their vector form.
Hence, only the contributions that include the velocity operator, which are shown in Eq.~(\ref{Electric-and-heat-current}), are relevant to longitudinal thermoelectric and thermal transport in a magnetic field.

\subsection{Theoretical model \label{Theoretical-model}}
We consider the continuum 3D Weyl Hamiltonian
\begin{align}
\mathcal{H}(\bm{k})=v_F(k_x\sigma_x+k_y\sigma_y)+m(k_z)\sigma_z,
\label{H_pm}
\end{align}
where $v_F$ is the Fermi velocity and $\sigma_i$ are the Pauli matrices.
For the simplest case (isotropic Weyl cone), we can set $m(k_z)=Qv_F k_z$ with $Q$ being the chirality of a given Weyl node.
For a two-node Weyl semimetal with broken time-reversal symmetry (which can be regarded as a system of a 3D topological insulator doped with magnetic impurities)  \cite{Burkov2011,Vazifeh2013,Sekine2014,Burkov2015,Sekine2017}, we can set $m(k_z)=b-\sqrt{v_F^2 k_z^2+\Delta^2}$, where $\Delta$ is the mass of 3D Dirac fermions describing the 3D topological insulator, and $b$ is the strength of a magnetic interaction such as $s$-$d$ coupling.
In this case, the two Weyl nodes are located on the $k_z$ axis as $W_\pm=(0,0,\pm k_0)$ with $k_0=\sqrt{b^2-\Delta^2}/v_F$.

The eigenvectors of the Hamiltonian~(\ref{H_pm}) with eigenvalues $\varepsilon^\pm_{\bm{k}}=\pm\varepsilon_{\bm{k}}=\pm\sqrt{v_F^2(k_x^2+k_y^2)+m^2}$ are given by
\begin{align}
|u_{\bm{k}}^\pm\rangle=\frac{1}{\sqrt{2}}
\begin{bmatrix}
\sqrt{1\pm\frac{m(k_z)}{\varepsilon_{\bm{k}}}}\\
\pm e^{i\theta}\sqrt{1\mp\frac{m(k_z)}{\varepsilon_{\bm{k}}}}
\end{bmatrix},
\label{WSM-eigenstates}
\end{align}
where $e^{i\theta}=(k_x+ik_y)/k_\perp$ with $k_\perp=\sqrt{k_x^2+k_y^2}$.
The generalized Berry connection in the eigenstate representation is given by $[\mathcal{R}_{\bm{k},\alpha}]^{mn}=i\langle u_{\bm{k}}^m|\partial_{k_\alpha}u_{\bm{k}}^n\rangle$ with $\alpha=x,y,z$ and $m,n=\pm$.
The individual components are given explicitly by
\begin{align}
\mathcal{R}_{\bm{k},x}
=&\ \frac{1}{2 k_\perp}\sin\theta-\tilde{\sigma}_z\frac{1}{2 k_\perp}\frac{m}{\varepsilon_{\bm{k}}}\sin\theta -\tilde{\sigma}_y\frac{v_F m}{2\varepsilon_{\bm{k}}^2}\cos\theta\nonumber\\
& -\tilde{\sigma}_x\frac{v_F}{2\varepsilon_{\bm{k}}}\sin\theta, \nonumber\\
\mathcal{R}_{\bm{k},y}
=&-\frac{1}{2 k_\perp}\cos\theta+\tilde{\sigma}_z\frac{1}{2 k_\perp}\frac{m}{\varepsilon_{\bm{k}}}\cos\theta -\tilde{\sigma}_y\frac{v_F m}{2\varepsilon_{\bm{k}}^2}\sin\theta \nonumber\\
& +\tilde{\sigma}_x\frac{v_F}{2\varepsilon_{\bm{k}}}\cos\theta, \nonumber\\
\mathcal{R}_{\bm{k},z}
=&\ \tilde{\sigma}_y\frac{v_F k_\perp}{2\varepsilon_{\bm{k}}^2}\frac{\partial m}{\partial k_z},
\label{Berry-connection}
\end{align}
where $\tilde{\sigma}_\alpha$ are the Pauli matrices in the eigenstate basis of
$\begin{bmatrix}
++ && +-\\
-+ && --
\end{bmatrix}$.
Also, the individual components of the Berry curvature, $\Omega^\pm_{\bm{k},a}=\epsilon^{abc}\, i\langle \partial_{k_b} u_{\bm{k}}^\pm|\partial_{k_c} u_{\bm{k}}^\pm\rangle$, are given by
\begin{align}
&\Omega^\pm_{\bm{k},x}=\mp\frac{\partial m}{\partial k_z}\frac{v_F^2 k_x}{2\varepsilon^3_{\bm{k}}},\ \ \ \ \ \Omega^\pm_{\bm{k},y}=\mp\frac{\partial m}{\partial k_z}\frac{v_F^2 k_y}{2\varepsilon^3_{\bm{k}}},\ \ \ \ \ \Omega^\pm_{\bm{k},z}=\mp\frac{v_F^2 m}{2\varepsilon^3_{\bm{k}}}.
\label{Berry-curvature}
\end{align}
Notice that the Berry curvature in Weyl semimetals has all the three ($x,y,z$) components, whereas the Berry curvature in 2D systems such as monolayer MoS$_2$ has only the out-of-plane component.

\subsection{Calculation of the density matrix}
\begin{figure*}[!t]
\centering
\includegraphics[width=2\columnwidth]{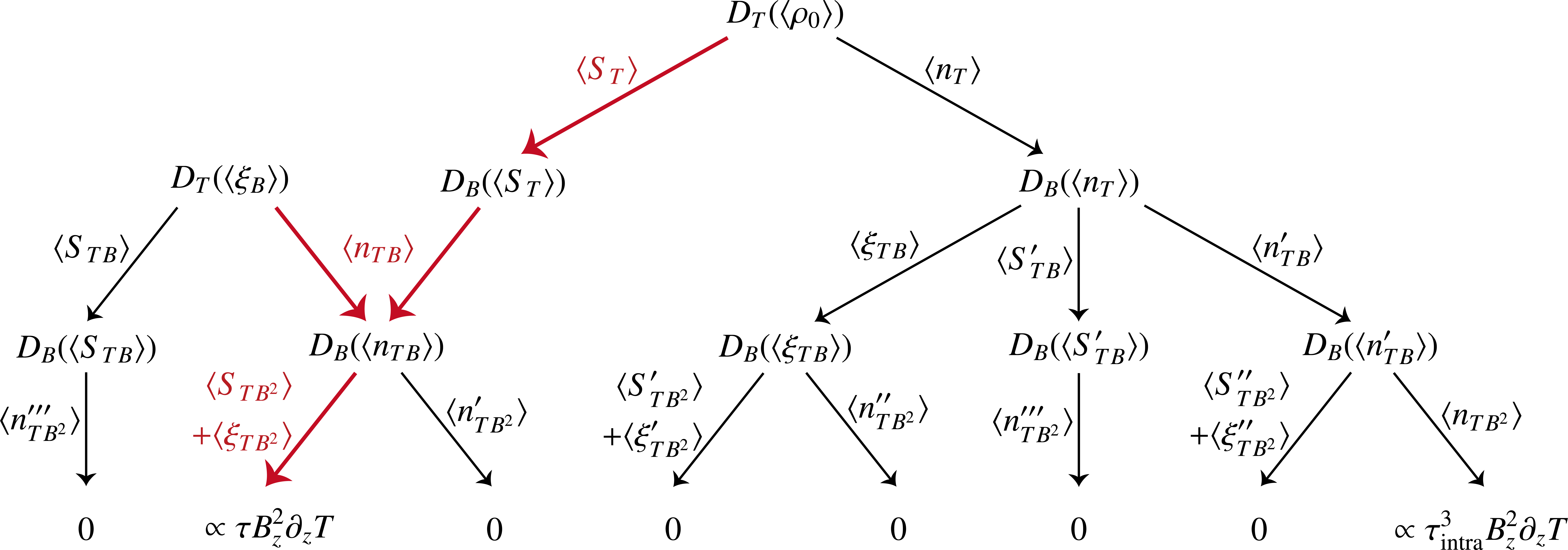}
\caption{Schematic illustration of the procedure for calculating the density matrix that is linear in temperature gradient and quadratic in magnetic field in a parallel configuration such that $\bm{E}_T=(0,0,-\partial_z T/T)$ and $\bm{B}=(0,0,B_z)$ for a Weyl semimetal described by the Hamiltonian~(\ref{H_pm}).
$\langle n\rangle$ and $\langle \xi\rangle$ indicates band-diagonal density matrix components, and $\langle S\rangle$ indicates band-off-diagonal density matrix components.
Prime marks indicate different contributions at each order of $B_z$.
Bold red arrows indicate the processes that contribute to the longitudinal thermoelectric and thermal conductivities induced by the thermal chiral anomaly.
The rightmost process that results in $\langle n_{TB^2}\rangle$ is purely extrinsic, i.e., indicates the contribution from the Lorentz force.
$\tau$ and $\tau_{\mathrm{intra}}$ indicate the intervalley and intravalley scattering times, respectively.
As in the case of the electric conductivity \cite{Sekine2017}, the other contributions to the longitudinal thermoelectric and thermal conductivities quadratic in magnetic field vanish.
}\label{Fig1}
\end{figure*}

Here, we briefly summarize our calculation of the density matrix $\langle\rho_{TB^2}\rangle$ that is linear in temperature gradient and quadratic in magnetic field.
In our theory, the formal expression for the density matrix $\langle\rho_{TB^2}\rangle$ in the low-field expansion is written from Eq.~(\ref{rho-expansion}) in the form
\begin{align}
\langle\rho_{TB^2}\rangle=(\mathcal{L}^{-1}D_B)^2 \mathcal{L}^{-1}D_T(\langle\rho_0\rangle)+\mathcal{L}^{-1}D_B \mathcal{L}^{-1}D_T(\langle\xi_B\rangle),
\label{rho_TB^2-formal}
\end{align}
where $\langle\rho_0\rangle$ is the Fermi-Dirac distribution function and $\langle\xi_B\rangle$ given by Eq.~(\ref{intrinsic-diagonal-matrix-B}).
Here, note that the angle between the temperature gradient and magnetic field is arbitrary in this formalism.
Since the purpose of this section is to investigate the relations of thermoelectric and heat currents to electric current (i.e., the Mott relation and the Wiedemann-Franz law, respectively), we follow the procedure for calculating the magnetoconductivity quadratic in magnetic field induced by the chiral anomaly \cite{Sekine2017} in order to obtain the thermoelectric and thermal conductivities quadratic in magnetic field.

We divide the calculation of the matrices $(\mathcal{L}^{-1}D_B)^2 \mathcal{L}^{-1}D_T(\langle\rho_0\rangle)$ and $\mathcal{L}^{-1}D_B \mathcal{L}^{-1}D_T(\langle\xi_B\rangle)$ in Eq.~(\ref{rho_TB^2-formal}) into four steps as follows.
First, we calculate the off-diagonal part (i.e., independent of disorder) of the density matrix induced by the temperature gradient,
\begin{align}
\langle S_T\rangle&=\mathcal{L}^{-1}D_T(\langle\rho_0\rangle)=[D_T(\langle\rho_0\rangle)]_{\bm{k}}^{mm'}/i(\varepsilon^m_{\bm{k}}-\varepsilon^{m'}_{\bm{k}}) \nonumber\\
&\propto \partial_z T,
\label{procedure1}
\end{align}
where $\langle\rho_0\rangle=\mathrm{diag}[f_0(\varepsilon^+_{\bm{k}}),f_0(\varepsilon^-_{\bm{k}})]$ with $f_0(\varepsilon^\pm_{\bm{k}})=1/[e^{(\pm \varepsilon_{\bm{k}}-\mu)/T}+1]$ being the Fermi-Dirac distribution function.
Second, we calculate the diagonal part of the density matrix that is linear in both temperature gradient and magnetic field, which results from $\langle S_T\rangle$ obtained in Eq.~(\ref{procedure1}) and $\langle \xi_B\rangle$,
\begin{align}
\langle n_{TB}\rangle&=\mathcal{L}^{-1}\left[D_B(\langle S_T\rangle)+D_T(\langle \xi_B\rangle)\right] \nonumber\\
&=\tau \left[D_B(\langle S_E\rangle)+D_E(\langle \xi_B\rangle)\right]_{\bm{k}}^{mm} \nonumber\\
&\propto e\partial_z T B_z\tau,
\label{procedure2}
\end{align}
where $\tau$ is the intervalley scattering time.
As we have seen in Eq.~(\ref{dN/dt-total}), the appearance of the intervalley scattering time $\tau$ is due to that the driving terms $D_B(\langle S_T\rangle)$ and $D_T(\langle \xi_B\rangle)$ have nonzero values when integrated over a given valley (Weyl cone).
In other words, this $\langle n_{TB}\rangle$ indeed arises as a consequence of the thermal chiral anomaly.
Third, we calculate the off-diagonal part of the density matrix that is linear in temperature gradient and quadratic in magnetic field, which results from $\langle n_{TB}\rangle$ obtained in Eq.~(\ref{procedure2}),
\begin{align}
\langle S_{TB^2}\rangle&=\mathcal{L}^{-1}D_B(\langle n_{TB}\rangle)=[D_B(\langle n_{TB}\rangle)]_{\bm{k}}^{mm'}/i(\varepsilon^m_{\bm{k}}-\varepsilon^{m'}_{\bm{k}}) \nonumber\\
&\propto e^2\partial_z T B_z^2\tau.
\label{procedure3}
\end{align}
Fourth, we calculate the intrinsic contribution to the diagonal part of the density matrix that is linear in temperature gradient and quadratic in magnetic field, which results from $\langle n_{TB}\rangle$ obtained in Eq.~(\ref{procedure2}),
\begin{align}
\langle \xi_{TB^2}\rangle&=P^{-1}D_B(\langle n_{TB}\rangle)=e\, \langle n_{TB}\rangle^{mm}_{\bm{k}}\bm{B}\cdot\bm{\Omega}^m_{\bm{k}} \nonumber\\
&\propto e^2\partial_z T B_z^2\tau.
\label{procedure4}
\end{align}
Figure.~\ref{Fig1} shows a schematic illustration of the procedure for calculating the density matrix that is linear in temperature gradient and quadratic in magnetic field.

In the following, we consider the low-temperature case where $T\ll \mu$, with $T$ and $\mu>0$ being the temperature and chemical potential of the system, respectively.
For the sake of clarity, we also consider the case of Weyl semimetals with isotropic Weyl cones, i.e., we set $m(k_z)=Qv_F k_z$.

\subsection{Thermoelectric conductivity \label{WSM-thermoelectric}}
Let us consider the case of a temperature gradient and a magnetic field in a parallel configuration such that $\bm{E}_T=(0,0,-\partial_z T/T)$ and $\bm{B}=(0,0,B_z)$.
We start by obtaining the off-diagonal part of the density matrix induced solely by the temperature gradient, $\langle  S_T\rangle$ [Eq.~(\ref{procedure1})].
Using the expressions for the thermal driving term $D_T(\langle\rho_0\rangle)$ [Eq.~(\ref{D_T-intrinsic})] and the Berry connection $\bm{\mathcal{R}}_{\bm{k}}$ [Eq.~(\ref{Berry-connection})], we get
\begin{align}
\langle  S_T\rangle=\tilde{\sigma}_y \frac{\partial_z T}{T}[(\varepsilon^+_{\bm{k}}-\mu)f_0(\varepsilon^+_{\bm{k}})-(\varepsilon^-_{\bm{k}}-\mu)f_0(\varepsilon^-_{\bm{k}})]\frac{v_F k_\perp}{4\varepsilon_{\bm{k}}^3}\frac{\partial m}{\partial k_z}.
\label{step1-explicit}
\end{align}
Here, note that we have not written down the contribution from $J(\langle n_T\rangle)$ in Eq.~(\ref{rho_T-off-diagonal}).
Such a contribution was shown to be zero for the magnetoconductivity in the case of short-range (on-site) disorder potential \cite{Sekine2017}.
Because the thermal driving term $D_T(\langle\rho_0\rangle)$ is quite similar to the electric driving term $D_E(\langle\rho_0\rangle)$ except for the factor $(\varepsilon^\pm_{\bm{k}}-\mu)$ in front of $f_0(\varepsilon^\pm_{\bm{k}})$, it turns out that the contribution from $J(\langle n_T\rangle)$ is also zero in the present case due to the fact that the factor $(\varepsilon^\pm_{\bm{k}}-\mu)$ is an even function of $\bm{k}$.

Second, we compute the diagonal density matrix $\langle n_{TB}\rangle$ proportional to $\partial_zT B_z$ [Eq.~(\ref{procedure2})].
This $\langle n_{TB}\rangle$ is the most important quantity in our formalism, since it directly reflects the electron-number nonconservation due to the thermal chiral anomaly, as shall be shown just below.
As described in Ref.~\cite{Sekine2017}, the magnetic driving term obtained from an off-diagonal matrix is purely diagonal for the Weyl Hamiltonian (\ref{H_pm}).
Following Ref.~\cite{Sekine2017} and using Eq.~(\ref{step1-explicit}), we have
\begin{align}
D_B(\langle S_T\rangle)&=\frac{eB_z}{2}\left[\left\{\frac{D\mathcal{H}}{Dk_y},\frac{D\langle S_T\rangle}{Dk_x}\right\}-\left\{\frac{D\mathcal{H}}{Dk_x},\frac{D\langle S_T\rangle}{Dk_y}\right\}\right]\nonumber\\
&=e \frac{\partial_z T}{T} B_z\mathcal{F}_{\bm{k}}\bm{1},
\label{step2-explicit}
\end{align}
where $\{\ \ ,\ \ \}$ indicates a matrix anticommutator.
Here,
\begin{align}
\mathcal{F}_{\bm{k}}=\ &-\frac{v_F^3 k_\perp}{\varepsilon_{\bm{k}}^2}c_{\bm{k}}-\frac{v_F m^2}{\varepsilon_{\bm{k}}^2k_\perp}c_{\bm{k}}-v_F\cos\theta\frac{\partial c_{\bm{k}}}{\partial k_x}-v_F\sin\theta\frac{\partial c_{\bm{k}}}{\partial k_y}\nonumber\\
=\ &\sum_{a=x,y}\left[\frac{1}{2}\Omega_{\bm{k},a}^+\frac{\partial}{\partial k_a} + 3k\, (\Omega_{\bm{k},a}^+)^2\right]\nonumber\\
&\times\left[(\varepsilon^+_{\bm{k}}-\mu)f_0(\varepsilon^+_{\bm{k}})-(\varepsilon^-_{\bm{k}}-\mu)f_0(\varepsilon^-_{\bm{k}})\right],
\label{A-B}
\end{align}
where $\bm{1}$ is the $2\times2$ identity matrix, $c_{\bm{k}}=[(\varepsilon^+_{\bm{k}}-\mu)f_0(\varepsilon^+_{\bm{k}})-(\varepsilon^-_{\bm{k}}-\mu)f_0(\varepsilon^-_{\bm{k}})](v_F k_\perp/4\varepsilon_{\bm{k}}^3)\partial m/\partial k_z$, and $k=\sqrt{k_x^2+k_y^2+k_z^2}$.
Also, using the expression for $\langle \xi_B\rangle$ [Eq.~(\ref{intrinsic-diagonal-matrix-B})], we have
\begin{align}
D_T(\langle \xi_B\rangle)=e \frac{\partial_z T}{T} B_z
\begin{bmatrix}
(\varepsilon^+_{\bm{k}}-\mu)\frac{\partial f_0(\varepsilon^+_{\bm{k}})\Omega^+_{\bm{k},z}}{\partial k_z} && 0\\
0 && (\varepsilon^-_{\bm{k}}-\mu)\frac{\partial f_0(\varepsilon^-_{\bm{k}})\Omega^-_{\bm{k},z}}{\partial k_z}
\end{bmatrix}.
\label{step2-explicit2}
\end{align}

Now, we show that these $D_B(\langle S_T\rangle)$ and $D_T(\langle \xi_B\rangle)$ have a special property.
We see that $\mathcal{F}_{\bm{k}}$ and $\partial [f_0(\varepsilon^{\pm}_{\bm{k}})\Omega^{\pm}_{\bm{k},z}]/\partial k_z$ are both even functions of $k_x$, $k_y$, and $k_z$.
Accordingly, we find that the integral of $D_B(\langle  S_T\rangle)+D_T(\langle \xi_B\rangle)$ over the Fermi surface of a given valley (Weyl cone) has a nonzero value: $\int_{\mathrm{FS}}\frac{d^3k}{(2\pi)^3}\, \sum_m [D_B(\langle  S_T\rangle)+D_T(\langle \xi_B\rangle)]_{\bm{k}}^{mm}\neq 0$.
As has been discussed in Sec.~\ref{Sec-CA-generic}, this is a consequence of the total electron number nonconservation in a given valley.
Namely, we obtain the rate of pumping of electrons between valleys
\begin{align}
\frac{\partial N}{\partial t}&=\frac{e B_z}{4\pi^2}\frac{\partial_z T}{T}\int_{\mathrm{FS}}\frac{d^3k}{2\pi}\left[2\, \mathcal{F}'_{\bm{k}}+\sum_{m=\pm}(\varepsilon^m_{\bm{k}}-\mu)\frac{\partial f_0(\varepsilon^m_{\bm{k}})}{\partial k_z}\Omega^m_{\bm{k},z}\right]\nonumber\\
&=\frac{e B_z}{4\pi^2}\partial_z T\sum_{m=\pm}\int \frac{d^3 \bm{k}}{2 \pi}\frac{\varepsilon^m_{\bm{k}}-\mu}{T}\frac{\partial f_{0}(\varepsilon^m_{\bm{k}})}{\partial \varepsilon^m_{\bm{k}}} \bm{v}^m_{\bm{k}} \cdot \bm{\Omega}^m_{\bm{k}},
\label{Nonzero-integrated-D_B}
\end{align}
which is indeed consistent with the general expression (\ref{dN/dt-total}).
Here, $\mathcal{F}'_{\bm{k}}$ is the component that represents the Fermi surface response, i.e., is proportional to $\partial f_0(\varepsilon^m_{\bm{k}})/\partial k_a$ ($a=x,y$) in Eq.~(\ref{A-B}).

Let us consider the consequence of the action of the scattering operator $\mathcal{L}^{-1}$ on $D_B(\langle S_T\rangle)$ and $D_T(\langle \xi_B\rangle)$.
In multivalley systems, the intervalley scattering time $\tau$ is in general much larger than the intravalley scattering time $\tau_{\mathrm{intra}}$ (i.e., $\tau_{\mathrm{intra}}/\tau\ll 1$), since the intervalley scattering processes require large momentum transfers, i.e., the number of intervalley scattering processes that can occur is much smaller than that of intravalley scattering processes.
Therefore, it follows that the intervalley scattering time $\tau$ appears as the largest eigenvalue of the matrix representation of $\mathcal{L}^{-1}$, when $\mathcal{L}^{-1}$ acts on $D_B(\langle S_T\rangle)$ and $D_T(\langle \xi_B\rangle)$.
(See Ref.~\cite{Sekine2017} for the detailed description of the properties of $\mathcal{L}^{-1}$.)
Then, the diagonal density matrix $\langle n_{EB}\rangle$ is obtained as
\begin{align}
\langle n_{TB}\rangle&=\mathcal{L}^{-1}\left[D_B(\langle S_T\rangle)+D_T(\langle \xi_B\rangle)\right]\nonumber\\
&=e^2 E_z B_z\tau
\begin{bmatrix}
\tilde{\mathcal{F}}_{\bm{k}}^{++} && 0\\
0 && \tilde{\mathcal{F}}_{\bm{k}}^{--}
\end{bmatrix},
\end{align}
where
\begin{align}
\tilde{\mathcal{F}}_{\bm{k}}^{mm}=&\ \frac{1}{2}\sum_{m'=\pm}(\varepsilon^{m'}_{\bm{k}}-\mu)\left[\frac{\partial f_0(\varepsilon^{m'}_{\bm{k}})}{\partial k_x}\Omega_{\bm{k},x}^{m'}+\frac{\partial f_0(\varepsilon^{m'}_{\bm{k}})}{\partial k_y}\Omega_{\bm{k},y}^{m'}\right]\nonumber\\
&+(\varepsilon^m_{\bm{k}}-\mu)\frac{\partial f_0(\varepsilon^m_{\bm{k}})}{\partial k_z}\Omega^m_{\bm{k},z}
\label{F_k-tilde}
\end{align}
is the component which represents the Fermi surface response in Eqs.~(\ref{step2-explicit}) and (\ref{step2-explicit2}).
Note that we have neglected the Fermi sea response in Eqs.~(\ref{step2-explicit}) and (\ref{step2-explicit2}).

Third, we compute the off-diagonal density matrix $\langle S_{TB^2}\rangle$ proportional to $\partial_zT B_z^2$ [Eq.~(\ref{procedure3})].
The off-diagonal part of the density matrix obtained from magnetic driving term acting on an arbitrary density matrix $\langle\rho\rangle$ is given by \cite{Sekine2017}
\begin{align}
\langle S_B\rangle_{\bm{k}}^{mm'}=-i\frac{[D_B(\langle\rho\rangle)]^{mm'}_{\bm{k}}-[J(\langle n\rangle)]^{mm'}_{\bm{k}}}{\varepsilon_{\bm{k}}^m-\varepsilon_{\bm{k}}^{m'}},
\label{rho_B-formal-expression}
\end{align}
where $m\neq m'$ and $\langle n\rangle$ is the off-diagonal part of $\langle\rho\rangle$.
Substituting $\langle\rho\rangle=\langle n_{TB}\rangle$ into Eq.~(\ref{rho_B-formal-expression}), we obtain the relevant off-diagonal density matrix as
\begin{align}
\langle S_{TB^2}\rangle=\ &\frac{e^2B_z^2\tau}{2}\frac{\partial_zT}{T} \nonumber\\
&\times\left[\frac{\partial (\tilde{\mathcal{F}}^{++}_{\bm{k}}+\tilde{\mathcal{F}}^{--}_{\bm{k}})}{\partial k_x}\left(\tilde{\sigma}_x\frac{v_F}{2\varepsilon_{\bm{k}}}\cos\theta-\tilde{\sigma}_y\frac{v_F m}{2\varepsilon_{\bm{k}}^2}\sin\theta\right)\right. \nonumber\\
&\left.+\frac{\partial (\tilde{\mathcal{F}}^{++}_{\bm{k}}+\tilde{\mathcal{F}}^{--}_{\bm{k}})}{\partial k_y}\left(\tilde{\sigma}_x\frac{v_F}{2\varepsilon_{\bm{k}}}\sin\theta+\tilde{\sigma}_y\frac{v_F m}{2\varepsilon_{\bm{k}}^2}\cos\theta\right)\right].
\label{step3-explicit}
\end{align}
Here, note that we have not written down the contribution from $J(\langle n_{TB}\rangle)$ in Eq.~(\ref{rho_B-formal-expression}), since it vanishes as in the case of the magnetoconductivity \cite{Sekine2017}.

Fourth, we compute the diagonal density matrix $\langle \xi_{TB^2}\rangle$ proportional to $\partial_zT B_z^2$ [Eq.~(\ref{procedure4})].
Notice that, in the presence of a magnetic field, there always exists the intrinsic Berry phase contribution to the diagonal part of a density matrix when $D_B$ acts on any band-diagonal density matrix \cite{Sekine2017} [see also Eq.~(\ref{intrinsic-diagonal-matrix-B})]:
\begin{align}
\langle \xi_B\rangle_{\bm{k}}^{mm}=[P^{-1}D_B(\langle n\rangle)]^{mm}_{\bm{k}}=e\, \mathcal{G}_{\bm{k}}^m\, \bm{B}\cdot\bm{\Omega}^m_{\bm{k}},
\label{intrinsic-diagonal-matrix-B-general}
\end{align}
where $\langle n\rangle^{mm'}_{\bm{k}}=\delta_{mm'}\mathcal{G}_{\bm{k}}^m$.
Substituting $\langle n\rangle=\langle n_{TB}\rangle$ into Eq.~(\ref{intrinsic-diagonal-matrix-B-general}), we readily obtain
\begin{align}
\langle\xi_{TB^2}\rangle=-e^2 \frac{\partial_z T}{T} B_z^2\tau \frac{v_F^2 m}{2\varepsilon^3_{\bm{k}}}
\begin{bmatrix}
\tilde{\mathcal{F}}_{\bm{k}}^{++} && 0\\
0 && -\tilde{\mathcal{F}}_{\bm{k}}^{--}
\end{bmatrix},
\label{step4-explicit-xi}
\end{align}
where we have used the explicit form of $\Omega_{z,\bm{k}}^{\pm}$ [Eq.~(\ref{Berry-curvature})].

We are now in a position to evaluate the $zz$-component of the chiral-anomaly induced thermoelectric conductivity proportional to $B_z^2$, which is given by $\alpha_{zz}^{\mathrm{CA}}=\mathrm{Tr}\{(-e)v_z[\langle S_{TB^2}\rangle+\langle \xi_{TB^2}\rangle]\}/(-\partial_zT)$.
Here, the velocity operator is written in the eigenstate basis as
\begin{align}
v_z=\frac{\partial m}{\partial k_z}\left(\frac{m}{\varepsilon_{\bm{k}}}\tilde{\sigma}_z+\frac{v_F k_\perp}{\varepsilon_{\bm{k}}}\tilde{\sigma}_x\right).
\label{v_z-matrix}
\end{align}
From Eqs.~(\ref{step3-explicit}), (\ref{step4-explicit-xi}), and (\ref{v_z-matrix}), an explicit expression for $\alpha_{zz}^{\mathrm{CA}}$ at low temperatures such that $T\ll \mu$ is obtained as
\begin{align}
\alpha_{zz}^{\mathrm{CA}}=\ &\frac{e^3B_z^2\tau}{2T}\int\frac{d^3k}{(2\pi)^3} \frac{\partial m}{\partial k_z}\frac{v_F^2}{\varepsilon_{\bm{k}}^2}\sum_{a=x,y}k_a\frac{\partial (\tilde{\mathcal{F}}^{++}_{\bm{k}}+\tilde{\mathcal{F}}^{--}_{\bm{k}})}{\partial k_a} \nonumber\\
& -\frac{e^3B_z^2\tau}{2T}\int\frac{d^3k}{(2\pi)^3} \frac{\partial m}{\partial k_z}\frac{v_F^2 m^2}{\varepsilon^4_{\bm{k}}}(\tilde{\mathcal{F}}^{++}_{\bm{k}}+\tilde{\mathcal{F}}^{--}_{\bm{k}})\nonumber\\
=& -\frac{e}{12}T\frac{(eB_z)^2v_F^3}{\mu^3}\tau,
\label{alpha-zz-CA}
\end{align}
where we have used $m(k_z)=Qv_F k_z$.
Note that Eq.~(\ref{alpha-zz-CA}) is the contribution from a given Weyl cone and it is independent of the chirality $Q$.
Therefore, the thermoelectric conductivity of a Weyl semimetal with $\mathcal{N}_{\mathrm{v}}$ nodes is given by $-\frac{e}{12}\mathcal{N}_{\mathrm{v}}T\frac{(eB_z)^2v_F^3}{\mu^3}\tau$.
We numerically find that $\alpha_{zz}^{\mathrm{CA}}$ is proportional to $T$, $1/\mu^3$, and $v_F^3$.
We also find that the Mott relation is satisfied as expected:
\begin{align}
\alpha_{zz}^{\mathrm{CA}}=\frac{\pi^2}{3e}T\left.\frac{\partial \sigma_{zz}^{\mathrm{CA}}}{\partial\mu}\right|_{T=0},
\end{align}
where $\sigma_{zz}^{\mathrm{CA}}=\frac{e^2}{8\pi^2}\frac{(eB_z)^2v_F^3}{\mu^2}\tau$ is the chiral-anomaly induced magnetoconductivity contributed from a given Weyl cone at zero temperature ($T=0$) \cite{Son2013,Burkov2014,Sekine2017}.

\subsection{Thermal conductivity}
Next, let us evaluate the $zz$-component of the chiral-anomaly induced thermal conductivity proportional to $B_z^2$, which is given by $\kappa_{zz}^{\mathrm{CA}}=\mathrm{Tr}\{\tfrac{1}{2}\{\mathcal{H}_0, v_z\}[\langle S_{TB^2}\rangle+\langle \xi_{TB^2}\rangle]\}/(-\partial_zT)$.
Here, the energy current operator is written in the eigenstate basis as
\begin{align}
\frac{1}{2}\{\mathcal{H}_0, v_z\}=\frac{\partial m}{\partial k_z}
\begin{bmatrix}
(\varepsilon^+_{\bm{k}}-\mu)\frac{m}{\varepsilon_{\bm{k}}} && -\mu\frac{v_F k_\perp}{\varepsilon_{\bm{k}}}\\
-\mu\frac{v_F k_\perp}{\varepsilon_{\bm{k}}} && -(\varepsilon^-_{\bm{k}}-\mu)\frac{m}{\varepsilon_{\bm{k}}}
\end{bmatrix}.
\label{H_0v_z-matrix}
\end{align}
Note that we have incorporated the chemical potential as $(\mathcal{H}_0)^{mn}=\delta_{mn}(\varepsilon^m_{\bm{k}}-\mu)$.
From Eqs.~(\ref{step3-explicit}), (\ref{step4-explicit-xi}), and (\ref{H_0v_z-matrix}), an explicit expression for $\kappa_{zz}^{\mathrm{CA}}$ at low temperatures such that $T\ll \mu$ is obtained as
\begin{align}
\kappa_{zz}^{\mathrm{CA}}=&-\frac{e^2B_z^2\tau}{2T}\int\frac{d^3k}{(2\pi)^3} \frac{\partial m}{\partial k_z}\frac{v_F^2}{\varepsilon_{\bm{k}}^2}(-\mu)\sum_{a=x,y}k_a\frac{\partial (\tilde{\mathcal{F}}^{++}_{\bm{k}}+\tilde{\mathcal{F}}^{--}_{\bm{k}})}{\partial k_a} \nonumber\\
& +\frac{e^2B_z^2\tau}{2T}\int\frac{d^3k}{(2\pi)^3} \frac{\partial m}{\partial k_z}\frac{v_F^2 m^2}{\varepsilon^4_{\bm{k}}}\sum_{n=\pm}(\varepsilon^n_{\bm{k}}-\mu)\tilde{\mathcal{F}}^{nn}_{\bm{k}}\nonumber\\
=\ & \frac{1}{24}T\mathcal{D}\frac{(eB_z)^2v_F^3}{\mu^2}\tau,
\label{kappa-zz-CA}
\end{align}
where we have used $m(k_z)=Qv_F k_z$ and $\mathcal{D}\approx 1.8$ is a dimensionless factor.
Note that Eq.~(\ref{kappa-zz-CA}) is the contribution from a given Weyl cone and it is independent of the chirality $Q$.
Therefore, the thermal conductivity of a Weyl semimetal with $\mathcal{N}_{\mathrm{v}}$ nodes is given by $\frac{1}{24}\mathcal{N}_{\mathrm{v}}T\mathcal{D}\frac{(eB_z)^2v_F^3}{\mu^2}\tau$.
We numerically find that $\kappa_{zz}^{\mathrm{CA}}$ is proportional to $T$, $1/\mu^2$, and $v_F^3$.
We also find that the Wiedemann-Franz law is violated due to the factor $\mathcal{D}$ as
\begin{align}
\kappa_{zz}^{\mathrm{CA}}=\frac{\pi^2}{3e^2}T\sigma_{zz}^{\mathrm{CA}}\times\mathcal{D}.
\end{align}
The violation of the Wiedemann-Franz law for the chiral-anomaly induced thermal conductivity has also been reported in recent theoretical studies \cite{Kim2014b,Lucas2016,Sharma2016,Nandy2019}.

It has been shown that the Wiedemann-Franz law should be retained in the presence of a quantizing magnetic field with $\omega_c\tau>1$ ($\omega_c$ is the cyclotron frequency and $\tau$ is the transport relaxation time) and if the condition $\omega_c\tau\gg k_BT$ is satisfied \cite{Smrcka1977}.
However, this is not the case in our study because we are focusing on the low magnetic field case with $\omega_c\tau\ll 1$ (i.e., when the Landau quantization can be neglected).
We expect that the Wiedemann-Franz law may be violated in systems with nontrivial Bloch bands in a low magnetic field, and that the violation of the Wiedemann-Franz law in a low magnetic field is due to a ``Berry phase effect''.

An experimentally observable value is the Lorenz number, which is defined by
\begin{align}
L(B_z)\equiv\frac{\kappa_{zz}}{T\sigma_{zz}}\approx\frac{1}{T}\frac{\kappa_{zz}^0+\kappa_{zz}^{\mathrm{CA}}}{\sigma_{zz}^0+\sigma_{zz}^{\mathrm{CA}}},
\label{Lorenz-number}
\end{align}
where $\sigma_{zz}$ and $\kappa_{zz}$ are the total electric and thermal conductivities, respectively.
$\sigma_{zz}^0$ and $\kappa_{zz}^0$ are the electric and thermal conductivities at zero magnetic field, respectively.
In Eq.~(\ref{Lorenz-number}) we have used the result that the contributions that are linear in magnetic field to $\sigma_{zz}$ and $\kappa_{zz}$ vanish in the configuration of $\nabla T \parallel \bm{B}$ and $\bm{E} \parallel \bm{B}$ \cite{Sekine2017}, and assumed that the contributions from the Lorentz force to $\sigma_{zz}$ and $\kappa_{zz}$ are small.
Since $\sigma_{zz}^{\mathrm{CA}}\ll \sigma_{zz}^0$ when the magnetic field is weak, we have
\begin{align}
\frac{\Delta L(B_z)}{L_0}\equiv\frac{L(B_z)-L_0}{L_0}
\approx(\mathcal{D}-1)\frac{\sigma_{zz}^{\mathrm{CA}}}{\sigma_{zz}^0}>0,
\end{align}
which is proportional to $B_z^2$.
Here, $L_0=\pi^2/3e^2$.
The enhancement of the Lorenz number has also been reported in a study invoking semiclassical wave-packet dynamics \cite{Sharma2016}.
It should be noted that the value of $\mathcal{D}$ depends on the detailed band structure of a system.

\section{Discussion \label{Sec-Discussion}}
So far we have developed a theory that describes the Berry phase effects on magnetotransport phenomena induced by a temperature gradient, utilizing a ``thermal vector potential'' theory \cite{Tatara2015}.
One may think that introducing a temperature gradient in the usual real-space partial derivative of a density matrix in Eq.~(\ref{full-kinetic-equation}) such that
\begin{align}
\frac{1}{2\hbar}\left\{\frac{D\mathcal{H}_0}{D\bm{k}} \cdot \nabla\langle \rho\rangle\right\}
\ \to\ \frac{1}{2\hbar}\left\{\frac{D\mathcal{H}_0}{D\bm{k}} \cdot \nabla T \frac{\partial\langle \rho\rangle}{\partial T}\right\}
\end{align}
works well instead of our formalism, i.e., without introducing the thermal driving term~(\ref{driving-term-T}).
However, we have checked that this way of introducing a temperature gradient does not correctly reproduce the anomalous Nernst effect [Eq.~(\ref{electric-current-general-expression})] and its counterpart due to the Onsager reciprocal relation [Eq.~(\ref{heat-current-general-expression})].
Our study implies that a temperature gradient should be regarded as a ``field'' that is described by a potential, as Luttinger originally proposed \cite{Luttinger1964}.
Although we expect that the same quantum kinetic equation as Eq.~(\ref{full-kinetic-equation}) can also be derived using the Luttinger's gravitational potential, it is not clear at present how to do it.

The Berry phase effects on electronic transport induced by a temperature gradient have usually been studied by combining semiclassical Boltzmann theory with semiclassical wave-packet dynamics, as in the case of the transport induced by an electric field.
In this semiclassical Boltzmann formalism, a temperature gradient is introduced by setting $\dot{\bm{r}}\cdot\frac{\partial}{\partial\bm{r}}\to \dot{\bm{r}}\cdot\frac{\partial T}{\partial\bm{r}}\frac{\partial}{\partial T}$ in the Boltzmann equation.
The calculation for obtaining physical observables becomes not straightforward when a magnetic field is present \cite{Kim2014a,Lundgren2014,Sharma2016,Nandy2019}: the form of the correction to the distribution function is assumed and then its solution is obtained by substituting the assumed form into the Boltzmann equation.
In contrast, the presence of a temperature gradient in our formalism is described as a driving force just as in the case of an electric field (i.e., comes from $\dot{\bm{k}}$ in the language of the semiclassical Boltzmann formalism).
Such an equivalence to an electric field makes it possible to systematically calculate the response of electron density matrices to a temperature gradient in powers of magnetic field.
This is an advantage of using our quantum kinetic formalism.
On the other hand, we note that our formalism is valid when the electron mean-free path is much longer than the electron wavelength, i.e., when electrons are weakly interacting, as in the case of the semiclassical Boltzmann theory.
Indeed, our quantum kinetic equation (\ref{full-kinetic-equation}) reduces to the semiclassical Boltzmann equation (\ref{semiclassical-Boltzmann}) in single-band systems without Berry curvatures.

The magnetic-field induced corrections in systems with nontrivial Bloch bands implied by semiclassical wave-packet dynamics are mainly divided into two: (1) the correction to the momentum-space density of states due to the Berry curvature, $(e/\hbar)\bm{B}\cdot\bm{\Omega}_{\bm{k}}^n$ \cite{Xiao2005,Xiao2010}, and (2) the correction to the Bloch electron energy due to the orbital magnetic moment, $-\bm{\mathfrak{m}}_{\bm{k}}^n\cdot\bm{B}$ \cite{Chang1996,Xiao2010}.
Here, $\bm{\Omega}_{\bm{k}}^n$ and $\bm{\mathfrak{m}}_{\bm{k}}^n$ are respectively the Berry curvature and orbital magnetic moment of a Bloch electron with momentum $\bm{k}$ in band $n$, and $\bm{B}$ is a magnetic field.
It has been shown that the correction to the momentum-space density of states due to the Berry curvature corresponds to a magnetic-field induced correction to the equilibrium single-particle density matrix in our quantum kinetic formalism \cite{Sekine2017,Sekine2018}.
However, it remains an important unresolved problem to derive the orbital magnetic moment itself and the resulting correction to the Bloch electron energy due to the orbital magnetic moment using our quantum kinetic formalism.

Our theory is applicable in principle to the calculation of the response of any single-particle observables to arbitrary order of the field strengths $E_i$, $B_j$, and $-\partial_k T$ for arbitrary field directions, as described in Eq.~(\ref{rho-arbitrary-order}).
All single-particle observables $\mathcal{O}$ maintain their crystal periodicity when they respond to spatially constant fields and therefore have expectation values of the form
\begin{align}
\langle \mathcal{O}\rangle=\mathrm{Tr}\left[\mathcal{O}\langle\rho\rangle\right],
\end{align}
where $\langle\rho\rangle$ is a density matrix we have derived in Eq.~(\ref{rho-arbitrary-order}).
The evaluation of field-induced spin currents and spin densities, which are related to the current-induced torques in spintronics, will be one of the practically important problems to which our quantum kinetic theory can be applied.
We anticipate, for example, that our theory will have interesting implications for the thermoelectric and thermal properties of 2D multivalley systems such as graphene and transition metal dichalcogenides, as well as 3D multivalley systems such as Weyl semimetals which we have studied in this paper.

\section{Summary  \label{Sec-Summary}}
In summary, we have developed a general quantum kinetic theory of thermoelectric and thermal transport in a low magnetic field that accounts for the interplay between momentum-space Berry curvatures, external electromagnetic fields and temperature gradient, and Bloch-state scattering.
The obtained quantum kinetic equation for Bloch states in the presence of disorder, electric and magnetic fields, and temperature gradient [Eq.~(\ref{full-kinetic-equation})], which is the principal result of this study, can be regarded as a matrix generalization of the usual semiclassical Boltzmann equation.
Our theory enables a systematic calculation of the linear and nonlinear responses of physical observables to temperature gradient in powers of magnetic field.
We have derived from a general Bloch Hamiltonian general expressions for the anomalous Nernst effect and its counterpart due to the Onsager reciprocal relation, which are in complete agreement with the expressions obtained by invoking semiclassical wave-packet dynamics.
However, the derivation of the anomalous thermal Hall effect remains a future subject.
We have also derived from a general Bloch Hamiltonian a general expression for the rate of pumping of electrons between valleys (i.e., the electron number nonconservation in a given valley due to the thermal chiral anomaly) in parallel temperature gradient and magnetic field.
From this expression we have found a relation, which is analogous to the Mott relation, between the rate of pumping due to a temperature gradient and that due to an electric field.
We have applied our theory to a simple two-band model for Weyl semimetals to study thermoelectric and thermal transport in a magnetic field.
We have shown that the Mott relation is satisfied in the chiral-anomaly induced thermoelectric conductivity, and that the Wiedemann-Franz law is violated in the chiral-anomaly induced thermal conductivity, which are both consistent with the results obtained by invoking semiclassical wave-packet dynamics.
In our derivation of the thermoelectric and thermal conductivities quadratic in magnetic field, the intervalley scattering time naturally appears as the eigenvalue of the scattering operator acting on the driving terms that are even functions of the momentum around a valley.

\acknowledgements
The authors thank Atsuo Shitade and Dimitrie Culcer for valuable discussions.
This work was supported by JST CREST Grant No. JPMJCR1874 and No. JPMJCR16F1, and by JSPS KAKENHI Grant No. 18H03676 and No. 26103006.
A.S. is supported by the Special Postdoctoral Researcher Program of RIKEN.

\appendix
\begin{widetext}
\section{General properties of the magnetic driving term $D_B$ \label{Appendix-D_B-1}}
In this Appendix, we consider general properties of the magnetic driving term (\ref{driving-term-B}).
We write the Hamiltonian of a system as $\mathcal{H}_0=\sum_{m}\varepsilon_m |m\rangle\langle m|$, where $\varepsilon_m$ is an energy eigenvalue of band $m$.
Note that we have omitted the wave vector $\bm{k}$ dependence to simplify the notation.
Without loss of generality we can set $\bm{B}=(0,0,B_z)$.
Let us consider the magnetic driving term acting on an arbitrary diagonal density matrix $\langle \mathcal{F}\rangle=\sum_m \mathcal{F}_{m} |m\rangle\langle m|$ with $m$ a band index.
The magnetic driving term (\ref{driving-term-B}) is written as
\begin{align}
D_B(\langle \mathcal{F}\rangle)=\frac{e}{2\hbar^2}\left\{\left(\frac{D \mathcal{H}_0}{D\bm{k}}\times\bm{B}\right)\cdot\frac{D\langle \mathcal{F}\rangle}{D\bm{k}}\right\}=\frac{eB_z}{2\hbar^2}\left[\left\{\frac{D\mathcal{H}_0}{Dk_y},\frac{D\langle \mathcal{F}\rangle}{Dk_x}\right\}-\left\{\frac{D\mathcal{H}_0}{Dk_x},\frac{D\langle \mathcal{F}\rangle}{Dk_y}\right\}\right].
\end{align}
First, we consider the diagonal component of $D_B(\langle \mathcal{F}\rangle)$, i.e., $\langle m|D_B(\langle\mathcal{F}\rangle)|m\rangle$.
We immediately get
\begin{align}
\frac{D\mathcal{H}_0}{Dk_y}&=\sum_{m'}\partial_y\varepsilon_{m'}|m'\rangle\langle m'|+\sum_{m'}\varepsilon_{m'}\bigl[|\partial_y m'\rangle\langle m'|+|m'\rangle\langle \partial_y m'|\bigr], \nonumber\\
\frac{D\langle \mathcal{F}\rangle}{Dk_x}&=\sum_{n'}\partial_x \mathcal{F}_{n'}|n'\rangle\langle n'|+\sum_{n'}\mathcal{F}_{n'}\bigl[|\partial_x n'\rangle\langle n'|+|n'\rangle\langle \partial_x n'|\bigr],
\label{appendix-D_B-element1}
\end{align}
where $\partial_a=\partial/\partial k_a$.
Then we have
\begin{align}
\frac{D\mathcal{H}_0}{Dk_y}\frac{D\langle \mathcal{F}\rangle}{Dk_x}=\ &\sum_{m'}\partial_y\varepsilon_{m'}\partial_x \mathcal{F}_{m'}|m'\rangle\langle m'|+\sum_{m' n'}\partial_y\varepsilon_{m'}\mathcal{F}_{n'}|m'\rangle\langle m'|\partial_x n'\rangle\langle n'|+\sum_{m'}\partial_y\varepsilon_{m'}\mathcal{F}_{m'}|m'\rangle\langle \partial_x m'| \nonumber\\
&+\sum_{m'}\varepsilon_{m'}\partial_x \mathcal{F}_{m'}|\partial_y m'\rangle\langle m'|+\sum_{m' n'}\varepsilon_{m'}\mathcal{F}_{n'}|\partial_y m'\rangle\langle m'|\partial_x n'\rangle\langle n'|+\sum_{m'}\varepsilon_{m'}\mathcal{F}_{m'}|\partial_y m'\rangle\langle \partial_x m'| \nonumber\\
&+\sum_{m' n'}\varepsilon_{m'}\partial_x \mathcal{F}_{n'}|m'\rangle\langle \partial_y m'|n'\rangle\langle n'|+\sum_{m' n'}\varepsilon_{m'}\mathcal{F}_{n'}|m'\rangle\langle \partial_y m'|\partial_x n'\rangle\langle n'|+\sum_{m' n'}\varepsilon_{m'}\mathcal{F}_{n'}|m'\rangle\langle \partial_y m'|n'\rangle\langle \partial_x n'|,
\label{appendix-D_B-element2}
\end{align}
from which the diagonal component is obtained as
\begin{align}
\langle m|\frac{D\mathcal{H}_0}{Dk_y}\frac{D\langle \mathcal{F}\rangle}{Dk_x}|m\rangle=\ &\partial_y\varepsilon_{m}\partial_x \mathcal{F}_{m}+\sum_{m'}\varepsilon_{m'}\mathcal{F}_{m}\langle m|\partial_y m'\rangle\langle m'|\partial_x m\rangle+\sum_{m'}\varepsilon_{m'}\mathcal{F}_{m'}\langle m|\partial_y m'\rangle\langle \partial_x m'| m\rangle \nonumber\\
&+\varepsilon_{m}\mathcal{F}_{m}\langle \partial_y m|\partial_x m\rangle+\sum_{n'}\varepsilon_{m}\mathcal{F}_{n'}\langle \partial_y m|n'\rangle\langle \partial_x n'| m\rangle,
\end{align}
where we have used that $\langle m'|\partial_a n'\rangle+\langle\partial_a  m'|n'\rangle=\partial_a (\delta_{m'n'})=0$.
Similarly, we have
\begin{align}
\langle m|\frac{D\langle \mathcal{F}\rangle}{Dk_x}\frac{D\mathcal{H}_0}{Dk_y}|m\rangle=\ &\partial_y\varepsilon_{m}\partial_x \mathcal{F}_{m}+\sum_{n'}\mathcal{F}_{n'}\varepsilon_{m}\langle m|\partial_x n'\rangle\langle n'|\partial_y m\rangle+\sum_{n'}\mathcal{F}_{n'}\varepsilon_{n'}\langle m|\partial_x n'\rangle\langle \partial_y n'| m\rangle \nonumber\\
&+\mathcal{F}_{m}\varepsilon_{m}\langle \partial_x m|\partial_y m\rangle+\sum_{m'}\mathcal{F}_{m}\varepsilon_{m'}\langle \partial_x m|m'\rangle\langle \partial_y m'| m\rangle.
\end{align}
FInally, we see that
\begin{align}
\langle m|D_B(\langle \mathcal{F}\rangle)|m\rangle=\frac{eB_z}{\hbar^2}(\partial_y\varepsilon_{m}\partial_x \mathcal{F}_{m}-\partial_x\varepsilon_{m}\partial_y \mathcal{F}_{m}).
\label{diagonal-element-D_B-diagonal-f}
\end{align}

\section{General expression for the rate of pumping $\partial N/\partial t$ in parallel temperature gradient and magnetic field \label{dN/dt-analytical}}
In this Appendix, we derive a general expression for the rate of pumping of electrons between valleys from a general microscopic electron model with the Hamiltonian $\mathcal{H}_0=\sum_{m}\varepsilon_m |m\rangle\langle m|$ and the equilibrium density matrix $\langle\rho_0\rangle=\sum_m f_{0m} |m\rangle\langle m|$, where $\varepsilon_m$ is an energy eigenvalue of band $m$ and $f_{0m}$ is the Fermi-Dirac distribution function of band $m$.
Note that we have omitted the wave vector $\bm{k}$ dependences in these equations to simplify the notation.
For concreteness, we consider the case of $\bm{E}_T=(0,0,-\partial_z T/T)$ and $\bm{B}=(0,0,B_z)$.
In Eq.~(\ref{S_T-general-expression}) the off-diagonal part of the density matrix induced by the temperature gradient has been obtained as
\begin{align}
\langle S_T\rangle=-i\frac{\partial_z T}{T}\sum_{nn'}\frac{\varepsilon_{n'}f_{0n'}-\varepsilon_{n}f_{0n}}{\varepsilon_n-\varepsilon_{n'}}|n\rangle\langle n|\partial_z n'\rangle\langle n'|,
\end{align}
where $n\neq n'$.
On the other hand, the magnetic driving term (\ref{driving-term-B}) acting on $\langle S_T\rangle$ is written as
\begin{align}
D_B(\langle S_T\rangle)=\frac{e}{2\hbar^2}\left\{\left(\frac{D \mathcal{H}_0}{D\bm{k}}\times\bm{B}\right)\cdot\frac{D\langle S_T\rangle}{D\bm{k}}\right\}
=\frac{eB_z}{2\hbar^2}\left[\left\{\frac{D\mathcal{H}_0}{Dk_y},\frac{D\langle S_T\rangle}{Dk_x}\right\}-\left\{\frac{D\mathcal{H}_0}{Dk_x},\frac{D\langle S_T\rangle}{Dk_y}\right\}\right].
\end{align}
Since we are focusing on the Fermi surface response, we consider only the terms proportional to $\partial_x f_{0}$ in $D\langle S_T\rangle/Dk_x$:
\begin{align}
\frac{D\langle S_T\rangle}{Dk_x}=-i\frac{\partial_z T}{T}\sum_{nn'}\frac{\varepsilon_{n'}\partial_xf_{0n'}-\varepsilon_{n}\partial_xf_{0n}}{\varepsilon_n-\varepsilon_{n'}}|n\rangle\langle n|\partial_z n'\rangle\langle n'|.
\end{align}
We also have the relevant term
\begin{align}
\frac{D\mathcal{H}_0}{Dk_y}=\frac{1}{\hbar}\sum_{m'}(\varepsilon_{m'}-\varepsilon_{n'})\bigl[|\partial_y m'\rangle\langle m'|+|m'\rangle\langle \partial_y m'|\bigr],
\end{align}
where we have used $\partial_a(\sum_{m'}|m'\rangle\langle m'|)=0$.
Note that the terms proportional to $\partial_y\varepsilon$ in $D\mathcal{H}_0/Dk_y$ do not contribute to the diagonal part of $D_B(\langle S_T\rangle)$.
Then, we obtain
\begin{align}
\langle m|\frac{D\mathcal{H}_0}{Dk_y}\frac{D\langle S_T\rangle}{Dk_x}|m\rangle
&=-\frac{i}{\hbar}\frac{\partial_z T}{T}\sum_{nn'}\sum_{m'}(\varepsilon_{m'}-\varepsilon_{n'})\frac{\varepsilon_{n'}\partial_xf_{0n'}-\varepsilon_{n}\partial_xf_{0n}}{\varepsilon_n-\varepsilon_{n'}}\delta_{n'm}\bigl[\delta_{m'n}\langle m|\partial_y n\rangle\langle n|\partial_z m\rangle+\delta_{mm'}\langle \partial_y m|n\rangle\langle n|\partial_z m\rangle\bigr] \nonumber\\
&=-\frac{i}{\hbar}\frac{\partial_z T}{T}\sum_{n}(\varepsilon_{m}\partial_x f_{0m}-\varepsilon_{n}\partial_x f_{0n})\langle m|\partial_y n\rangle\langle n|\partial_z m\rangle \nonumber\\
&=\frac{i}{\hbar}\frac{\partial_z T}{T}\varepsilon_{m}\partial_x f_{0m}\langle \partial_y m|\partial_z m\rangle-\frac{i}{\hbar}\frac{\partial_z T}{T}\sum_n\varepsilon_{n}\partial_x f_{0n}\langle\partial_z  n|m\rangle\langle m|\partial_y n\rangle.
\end{align}
Similarly, we have
\begin{align}
\langle m|\frac{D\langle S_T\rangle}{Dk_x}\frac{D\mathcal{H}_0}{Dk_y}|m\rangle=-\frac{i}{\hbar}\frac{\partial_z T}{T}\varepsilon_{m}\partial_x f_{0m}\langle \partial_z m|\partial_y m\rangle+\frac{i}{\hbar}\frac{\partial_z T}{T}\sum_n\varepsilon_{n}\partial_x f_{0n}\langle\partial_y  n|m\rangle\langle m|\partial_z n\rangle.
\end{align}
Finally, we arrive at a general expression for the rate of pumping from the Fermi surface contribution:
\begin{align}
\mathrm{Tr}\left[D_B(\langle S_T\rangle)\right]=\frac{eB_z}{\hbar^3}\frac{\partial_z T}{T}\sum_{m,\bm{k}}\varepsilon_{m}\left[\partial_x f_{0m}\Omega_{x}^m+\partial_y f_{0m}\Omega_{y}^m\right],
\end{align}
where $\Omega_{a}^m=\epsilon^{abc}\, i\langle \partial_{b} m|\partial_{c} m\rangle$ is the Berry curvature of band $m$.

\end{widetext}


\end{document}